\documentclass[journal=nalefd,manuscript=letter,layout=onecolumn]{achemso}
\pdfoutput=1
\usepackage{graphicx}
\usepackage[utf8]{inputenc}
\usepackage{amsmath}
\usepackage{braket}
\usepackage{lmodern}
\usepackage[normalem]{ulem}
\usepackage{bbold}
\usepackage[T1]{fontenc}
\usepackage{xcolor}
\usepackage{bm}
\usepackage{siunitx}
\usepackage{hyperref}
\hypersetup{pdfnewwindow=true,allcolors=black,colorlinks=true}
\usepackage{lineno}
\usepackage{here}

\let\s\textsubscript
\let\vec\bm

\newcommand{\nbs}{NbS\s2}

\def\UK{%
	II. Physikalisches Institut,
	Universit\"at zu K\"oln,
	Z\"ulpicher Stra\ss e 77,
	D-50937 K\"oln,
	Germany}

\def\UHB{%
	U~Bremen Excellence Chair,
	Bremen Center for Computational Materials Science,
	and MAPEX Center for Materials and Processes,
	University of Bremen,
	D-28359 Bremen,
	Germany}

\def\UHH{%
	I. Institute of Theoretical Physics,
	Universit\"at Hamburg,
	D-22607 Hamburg,
	Germany}

\def\CUIHH{%
	The Hamburg Centre for Ultrafast Imaging,
	Luruper Chaussee 149,
	D-22761 Hamburg,
	Germany}

\def\LU{%
	Division of Mathematical Physics,
	Department of Physics,
	Lund University,
	SE-22100 Lund,
	Sweden}

\title{Unconventional charge-density-wave gap in monolayer \nbs}

\author{Timo~Knispel}
\affiliation\UK

\author{Jan~Berges}
\affiliation\UHB

\author{Arne~Schobert}
\affiliation\UHH

\author{Erik~G.~C.~P.~van~Loon}
\affiliation\LU

\author{Wouter~Jolie}
\affiliation\UK

\author{Tim~Wehling}
\affiliation\UHH
\alsoaffiliation\CUIHH

\author{Thomas~Michely}
\affiliation\UK

\author{Jeison~Fischer}
\email{jfischer@ph2.uni-koeln.de}
\affiliation\UK

\date{\today}

\begin{document}
\DeclareGraphicsExtensions{.pdf}

\begin{tocentry}
\includegraphics{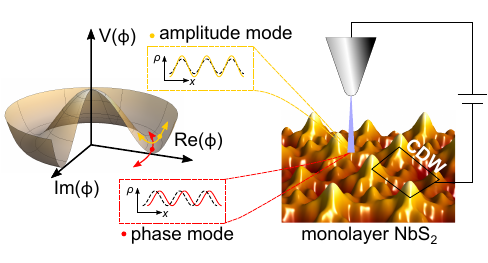}
\end{tocentry}

\newpage
\begin{abstract}
Using scanning tunneling microscopy and spectroscopy, for a monolayer of transition metal dichalcogenide H-\nbs~grown by molecular beam epitaxy on graphene, we provide unambiguous evidence for a charge density wave (CDW) with a $3\times3$ superstructure, which is not present in bulk \nbs. Local spectroscopy displays a pronounced gap of the order of 20\,meV at the Fermi level. Within the gap low energy features are present. The gap structure with its low energy features is at variance with the expectation for a gap opening in the electronic band structure due to a CDW. Instead, comparison with \emph{ab initio} calculations indicates that the observed gap structure must be attributed to combined electron-phonon quasiparticles. The phonons in question are the elusive amplitude and phase collective modes of the CDW transition. Our findings advance the understanding of CDW mechanisms in two dimensional materials and their spectroscopic signatures.
\end{abstract}
\vspace{3mm}
\hrule
\newpage

Interacting electron systems give rise to a diverse array of ordered states at low temperatures, such as superconductivity~\cite{qiu_SC_2021}, magnetism~\cite{mak_magnetism_2019}, and charge density waves (CDWs)~\cite{Rossnagel11}. These ordering tendencies generically stem from the interplay of kinetic and interaction energies with entropy. The ordering-induced energy gains often translate into the opening of gaps in the electronic excitation spectra~\cite{slater_magnetic_1951,Peierls55,bardeen_microscopic_1957,BCS_Long_1957}. Spectroscopy of these electronic gaps has been instrumental in understanding the nature of these ordered states: The momentum structure of the gap as well as its response to impurities point toward order-parameter symmetries. Typically, the comparison of gaps with transition temperatures is instrumental in discerning weak versus strong coupling physics, \emph{i.e.}, to distinguish between the BCS (Bardeen-Cooper-Schrieffer) and BEC (Bose-Einstein condensate) limits of superconductivity~\cite{Chen05}, between Slater and Heisenberg antiferromagnets~\cite{Schafer15}, or between Peierls to strong coupling CDWs~\cite{Rossnagel11}. Time-dependent gap spectroscopy in pump-probe setups~\cite{Demsar99,Petersen11,Liu13} offers a means to identify the relevant degrees of freedom associated with a certain type of order. Correspondingly, the analysis of gaps has been widely used to pinpoint which mechanism is responsible for CDW formation.

A well-established CDW mechanism is Fermi surface nesting in the classical Peierls picture~\cite{Peierls55}, valid for a one dimensional metallic chain developing a periodic lattice distortion. Due to the electronic response of the system, such a distortion is accompanied by an energy gap emerging at the Fermi energy and charge modulation with its periodicity given by the so-called nesting wave vector. CDWs in real materials can deviate from this idealized Peierls picture in several ways. For many materials, the electron-phonon coupling is strongly wave-vector dependent, which becomes the force driving the CDW~\cite{Johannes08,Weber11,Berges20}. Instead of gapping out the entire Fermi surface, the wave-vector-dependent electron-phonon coupling can induce partial gaps and changes in the Fermi surface topology~\cite{vanEfferen21}. Generally, the change in the electronic structure is not limited to the Fermi energy, like in the classical Peierls transition, but occurs over a broader energy range~\cite{Rossnagel11,Hofmann19,vanEfferen21} --- particularly in strong coupling situations~\cite{Rossnagel11}. Regardless, however, of strong versus weak coupling scenarios, CDW gap opening is generically explained in a Born-Oppenheimer picture, where electrons move within an effectively static distorted lattice.

Here, we investigate the CDW in monolayer \nbs~using scanning tunneling microscopy (STM) and spectroscopy (STS) as well as theoretical \emph{ab initio} based modeling. Within a clear gap in the STS measured d$I$/d$V$ spectra, a persistent and position dependent fine structure is observed. We demonstrate that the measured unconventional gap with its low energy spectral features reflects the robust presence of collective CDW phonon modes, specifically amplitude and phase modes, which couple to the electrons rather than the opening of a gap in the electronic band structure.

\begin{figure}[t]
	\centering
	\includegraphics[width=0.35\textwidth]{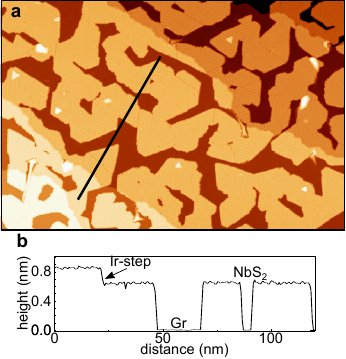}
	\caption{Structure of the \nbs~monolayer:
	(a)~STM topograph of coalesced monolayer islands of H-\nbs~on Gr/Ir(111).
	(b)~STM height profile along the black line in (a).
	Image information:
	(a)~size $\mathrm{250\,nm \times 166\,nm}$, $V_\mathrm{s} = 1$\,V, $I_\mathrm{t} = 0.1$\,nA, $T_\mathrm{s} = 1.7$\,K.
	}
	\label{fig_topo}
\end{figure}

\begin{figure*}[t]
	\centering
	\includegraphics[width=0.9\textwidth]{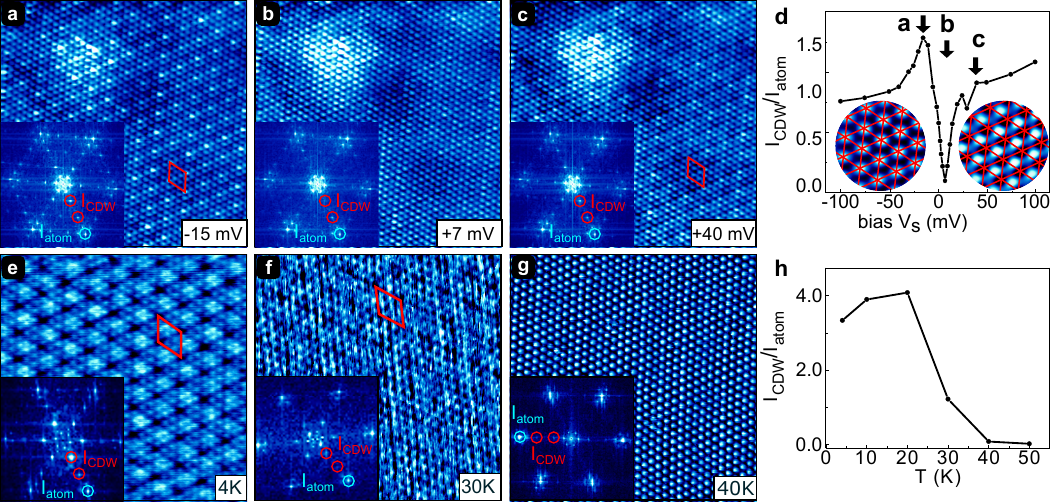}
	\caption{CDW in monolayer \nbs: (a), (b), and (c) are bandstop filtered constant-height d$I$/d$V$ maps taken at $V_\mathrm{s} = - 15$\,mV, $V_\mathrm{s} = 7$\,mV, and $V_\mathrm{s} = 40$\,mV, respectively. Red diamonds in (a) and (c) mark the $3\times3$ superstructure. The insets are the FFTs of the d$I$/d$V$ maps. Spots of the $3\times3$ superstructure are marked in red. (d) Intensity ratio $I_\mathrm{3\times3}/I_\mathrm{atom}$ of the first order $3\times3$ superstructure spots and first order \nbs~spots visible in the FFTs as a function of sample bias. The data points for (a), (b), and (c) are indicated. Insets of (d) highlight the lateral shift of superstructure maxima when crossing the Fermi level. The red line pattern is located at the exact same position in relation to the atomic lattice. Only first order $3\times3$ spots were back-transformed (see Figure~S4 of SI for details on FFT filtering).
	(e--g)~Constant-current d$I$/d$V$ maps taken at sample temperatures $T_\mathrm{s}$ indicated. The insets are the FFTs of the d$I$/d$V$ maps. Spots of the $3\times3$ superstructure are marked in red. (h) Intensity ratio $I_\mathrm{3\times3}/I_\mathrm{atom}$ as a function of temperature.
	Image information:
	(a--c)~size $\mathrm{12\,nm\times12}$\,nm, $V_\mathrm{stab} = 300$\,mV, $I_\mathrm{stab} =5$\,nA, $V_\mathrm{mod} = 5$\,mV, $f_\mathrm{mod} = 1890$\,Hz, $T_\mathrm{s} = 4$\,K, FFT filtered;
	(e--g)~size $\mathrm{9\,nm\times9}$\,nm, $V_\mathrm{s} = 100$\,mV, $I_\mathrm{t} = 0.7$\,nA, $V_\mathrm{mod} = 10$\,mV, $f_\mathrm{mod} = 1890$\,Hz.
	}
	\label{fig_cdw}
\end{figure*}

Monolayer \nbs~was grown \emph{in situ} on single crystal graphene (Gr) on Ir(111) and transferred in ultrahigh vacuum to the STM, see Supporting Information (SI) for details. The STM topograph of Figure~\ref{fig_topo}a displays coalesced monolayer islands covering most of the Gr substrate. (see Figure~S1 of SI for a low-energy electron diffraction pattern). The \nbs~layer conforms to Gr, which itself is continuous over Ir(111) steps running from the upper left to the lower right. The apparent height of \nbs~is in the range of 0.53--0.64\,nm depending on the tunneling parameters. An exemplary profile is shown in Figure~\ref{fig_topo}b taken along the black line of Figure~\ref{fig_topo}a. The atomic lattice of \nbs~as visible in all STM d$I$/d$V$ maps of Figure~\ref{fig_cdw} has a periodicity of 0.331(3)\,nm as measured by STM and low energy electron diffraction. The STS inferred density of states (DOS) in the range of $\pm$2\,eV around the Fermi energy (see Figure~S2 of SI) and the dispersion of the $\Gamma$~pocket measured by STS quasiparticle interference (see Figure~S3 of SI) make plain that the monolayer has H-\nbs~and not T-\nbs~structure. The measured in-plane lattice parameter and apparent height match with bulk values for 2H-\nbs~\cite{Fisher80} and previously measured monolayer values on Gr/SiC(0001)~\cite{Lin18} and Au(111)~\cite{Stan19}.

The sequence of constant-height fast Fourier transform (FFT) filtered d$I$/d$V$ maps in Figure~\ref{fig_cdw}a-c are all taken in the same area and with the same STM tip at sample bias voltages $V_\mathrm{s} = - 15$\,mV, $V_\mathrm{s} = 7$\,mV, and $V_\mathrm{s} = 40$\,mV, respectively (see Figure~S4 of SI for details on FFT filtering). While the atomically resolved maps in Figure~\ref{fig_cdw}a and c display a clear $3\times3$ superstructure, it is absent in Figure~\ref{fig_cdw}b, as expected for a map taken within a CDW gap. The intensity ratio $I_\mathrm{3\times3}/I_\mathrm{atom}$\cite{Ugeda16} of the first order $3\times3$ superstructure spots and the first order \nbs~lattice spots is shown as a function of bias voltage $V_\mathrm{s}$ in Figure~\ref{fig_cdw}d. The plot displays a clear minimum at $7$\,mV, where the intensity ratio is lower by a factor of 20 compared to the maximum at about $-15$\,mV. The maxima of the $3\times3$ superstructure shift between the d$I$/d$V$ maps taken at $V_\mathrm{s} = - 15$\,mV and at $V_\mathrm{s} = 40$\,mV as expected for a CDW when crossing its gap and as highlighted by the insets of Figure~\ref{fig_cdw}d.

Additional insight into the CDW stems from the temperature dependence of the $3\times3$ superstructure. Figure~\ref{fig_cdw}e--g shows a sequence of d$I$/d$V$ maps measured at sample temperatures of 4\,K, 30\,K, and 40\,K respectively. The superstructure intensity diminishes with increasing temperature and vanishes at 40\,K, as obvious from the topographs and their FFT insets. Figure~\ref{fig_cdw}h is a plot of $I_\mathrm{3\times3}/I_\mathrm{atom}$ as a function of temperature. Based on the data, we estimate a transition temperature $T_\mathrm{CDW} \approx 40$\,K.

Taken together, the existence of a $3\times3$ superstructure, the $I_\mathrm{3\times3}/I_\mathrm{atom}$ intensity ratio minimum next to the Fermi level, the phase shift of the superstructure when crossing the Fermi level, and the disappearance of the superstructure at 40\,K sum up to sound evidence for the presence of a CDW in monolayer \nbs.

Although unambiguous experimental evidence for a CDW in monolayer \nbs\ was missing, our finding is no surprise. While it is well established that bulk \nbs\ does not display a CDW~\cite{Naito82, Guillamon08}, it was pointed out that bulk \nbs\ is at the verge of forming a CDW due to strong electron-phonon coupling~\cite{Heil17}. In monolayer \nbs~ on Au(111) no CDW was observed~\cite{Stan19}, while on Gr on SiC(0001) the $3\times3$ superstructure was observed and attributed to a CDW~\cite{Lin18}. In subsequent theoretical investigations and using the experimental lattice parameter, the monolayer phonon dispersion indeed was shown to become unstable close to $q_\text{CDW} = 2/3\,\overline{\Gamma \textrm M}$~\cite{Bianco19}, which is indeed the CDW wave vector corresponding to the $3\times3$ superstructure observed.

\begin{figure*}[t]
	\centering
	\includegraphics[width=0.9\textwidth]{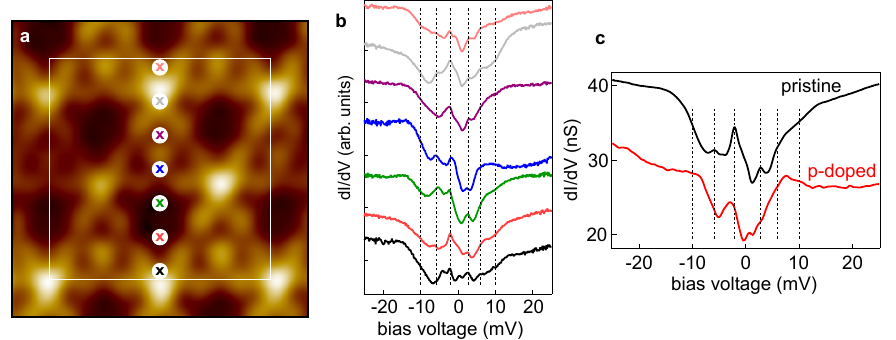}
	\caption{d$I$/d$V$ spectra of monolayer \nbs~near the Fermi-level:
	(a)~d$I$/d$V$ map of monolayer \nbs.
	(b)~High-resolution d$I$/d$V$ spectra taken at the points color-coded in (a). Thin dashed lines at $\pm10$\,mV highlight flanks of gap. Thin dashed lines highlight positions of toggling peaks within gap. The d$I$/d$V$ spectra are shifted vertically for better visualization.
	(c)~Black curve: average of $43\times43$ d$I$/d$V$ spectra taken within the white box in (a). Red curve: d$I$/d$V$ spectrum taken of monolayer \nbs~grown on O-intercalated graphene on Ir(111). The red d$I$/d$V$ spectrum is vertically shifted down by 5\,nS.
	Image information:
	(a)~size $\mathrm{2.7\,nm\times2.7}$\,nm, $V_\mathrm{s} = 100$\,mV, $I_\mathrm{t} = 1.0$\,nA, $V_\mathrm{mod} = 15$\,mV, $f_\mathrm{mod} = 1873$\,Hz, $T_\mathrm{s} = 0.4$\,K;
	(b)~$V_\mathrm{stab} = 100$\,mV, $I_\mathrm{stab} = 4.7$\,nA, $V_\mathrm{mod} = 0.5$\,mV, $f_\mathrm{mod} = 311$\,Hz, $T_\mathrm{s} = 0.4$\,K;
	(c)~black curve: $V_\mathrm{stab} = 100$\,mV, $I_\mathrm{stab} = 4.7$\,nA, $V_\mathrm{mod} = 0.5$\,mV, $f_\mathrm{mod} = 311$\,Hz, $T_\mathrm{s} = 0.4$\,K; red curve: $V_\mathrm{stab} = 50$\,mV, $I_\mathrm{stab} = 0.5$\,nA, $V_\mathrm{mod} = 0.5$\,mV, $f_\mathrm{mod} = 797$\,Hz, $T_\mathrm{s} = 0.4$\,K.
	}
	\label{fig_sts}
\end{figure*}

High-resolution d$I$/d$V$ spectra are taken along a high symmetry line in the $3\times3$ unit cell of the CDW at locations indicated in the d$I$/d$V$ map of Figure~\ref{fig_sts}a and represented in Figure~\ref{fig_sts}b. In all spectra at roughly $\pm10$\,mV (thin dashed lines) the d$I$/d$V$ intensity slopes down forming a trough valley with d$I$/d$V$ intensity reduced by 20--30\,\% (compare Figure~\ref{fig_sts}c). Inside the trough valley small peaks of d$I$/d$V$ intensity are visible. Despite a strong intensity variation of these peaks, if present, they tend to be at the same bias symmetric locations of $\pm6$\,mV and $\pm2.5$\,mV with a spread of less than 0.5\,meV. These locations are highlighted by dashed lines in Figure~\ref{fig_sts}b. Figure~\ref{fig_sts}c shows as black curve the average of $43\times43$ d$I$/d$V$ spectra taken within the white box in the image of Figure~\ref{fig_sts}a. The flanks of the trough valley are well visible, as are three out of the four inner peaks, while the fourth at $+6$\,mV appears as a shoulder. Figure~\ref{fig_sts}c also presents as red curve a d$I$/d$V$ spectrum of \nbs~with less electrons in the band structure, \emph{i.e.}, on p-doped \nbs. P-doping was achieved by oxygen intercalation under Gr (see Figure~S5 of SI for details), thereby increasing its work functions by around 0.5\,eV. Vacuum level alignment implies the build-up of an interface dipole through transfer of electrons out of \nbs~\cite{vanEfferen22}. Comparing the two spectra in Figure~\ref{fig_sts}c makes plain that the width of the trough valley decreased and the peaks at the bottom of the valley changed their position.

While a gap in the measured d$I$/d$V$ spectra located at the Fermi level is often taken as an indication of a gap in the electronic band structure, a CDW gap at the Fermi level is not necessarily reflected in a gap in an STS d$I$/d$V$ spectrum~\cite{vanEfferen21}. Although due to a CDW at least partial electronic band gaps will open at the Fermi level, they may be inconspicuous to STS. STS is mostly sensitive to electronic states with small parallel momentum. \nbs~has no states with small parallel momentum near the Fermi edge (compare Figure~S3 of SI).

Indeed, the trough shape of our gap does not appear like a single gap in the spectral function~\cite{Ryu17,Wan22}, but is more reminiscent to a spectrum resulting from an inelastic tunneling process setting in at about $\pm 10$\,meV~\cite{Zhang08,Gawronski08}. Through the additional tunneling channel the overall tunneling probability increases beyond the onset energy. One might be tempted to assign the inelastic feature at $\pm 10$\,mV to bulk phonon modes of \nbs~expected at $\pm12$\,mV~\cite{Nishio94}. Such modes have been observed in STS of defected bulk 2H-\nbs~\cite{Wen20} and bulk 2H-NbSe$_2$\cite{Hou20}, but displayed no link to a CDW. In addition, the substantial reduction of the gap and its changed internal features upon p-doping rule out this assumption, as bulk phonon modes are not expected to change drastically upon doping.

Worse yet, none of the ideas invoked up to now provide an explanation for the internal fine structure of our gap with tiny peaks at $\pm 2.5$\,meV and $\pm 6$\,meV. However, strong indication that these features are related to the CDW is given by the spatial distribution of the peaks, that retains the periodicity of the CDW (see Figure~S6 in the SI).

We note that the gap and its internal peak structure in the d$I$/d$V$ spectra are unchanged through an external magnetic field of up to 8\,T. This rules out a superconducting or magnetic origin (see Figure~S7 of SI), such as the spin density waves which have been discussed in the theory literature~\cite{Xu14,Guller16,vanloon18}.

In search for an explanation for the observed features in the low-energy d$I$/d$V$ spectra, we perform calculations based on density functional theory (DFT) and density functional perturbation theory (DFPT), which provide us with the electronic and phononic properties, respectively. Since DFT and especially DFPT for large systems are computationally costly, we use the downfolding strategy described in Ref.~\citenum{Schobert2023} to construct a low-energy model from a single calculation in the undistorted phase. Within this downfolded model, we can then efficiently calculate the (free) energy, forces, and electronic band structure in the distorted phase, which requires a supercell that is several times larger than the original unit cell. Here, we briefly remark on the Marzari-Vanderbilt cold smearing~\cite{Marzari1999} parameter $\sigma$, which is used to stabilize the simulation of metals. Varying this parameter illustrates how stable the results are and acts as an indication of the effects of temperature and substrate hybridization~\cite{Hall19} (see SI for more details).

\begin{figure*}
	\includegraphics[width=\linewidth]{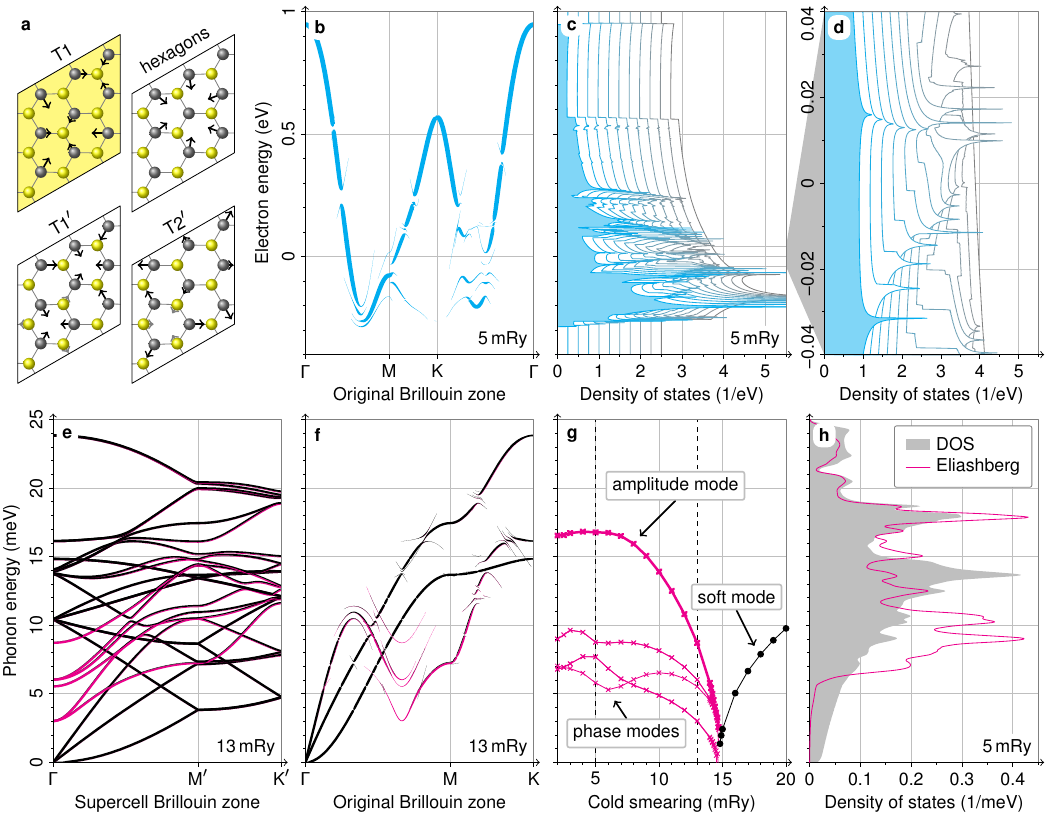}
	\caption{DFT-based low-energy model of \nbs:
	(a)~Four stable CDW structures. Atomic positions are drawn to scale, Nb in gray and S in yellow. Arrows indicating atomic displacements are scaled by a factor of 15 for visibility. The T1 structure where Nb atoms move toward S atoms is considered in the following (highlighted).
	(b)~Unfolded electronic band structure of the T1 structure. The Fermi level is located at zero.
	(c,\,d)~Corresponding DOS for CDW displacements scaled by a factor $\alpha$ between 0 (gray, rightmost curve) and 1 (blue, leftmost curve). (d) is a close-up of the low-energy region.
	(e,\,f)~Phonon dispersion just below the CDW transition. Displacement components belonging to the phase and amplitude modes are shown in magenta. Here, the colored fraction of the linewidth represents the summed squared modulus of the eigenvector overlap. The dispersion in (f) has been unfolded to the Brillouin zone of the symmetric structure.
	(g)~Transition of the softened phonon mode into amplitude (thick line) and phase modes.
	(h)~Phonon DOS and dimensionless (normalized) Eliashberg spectral function. A smearing parameter of 5 mRy is used in (b-d,\,h), while 13 mRy is used in (e,\,f). These values are indicated as vertical lines in (g).}
	\label{fig:theory}
\end{figure*}

The experimentally observed CDW phase involves a distortion of the original atomic lattice into a $3\times 3$ superstructure. A DFPT calculation of the phonon spectrum in the symmetric (undistorted) phase shows several degenerate unstable phonon modes that would result in a $3\times 3$ superstructure. By relaxation of a $3 \times 3$ supercell starting from randomized atomic positions within the model, we were able to identify four qualitatively different possible CDW structures, shown in Figure~\ref{fig:theory}\,(a). They all feature a significant displacement of the Nb atoms that preserves both the in-plane mirror symmetry and the three-fold rotation symmetry at the points toward or away from which the Nb atoms move. As in Ref.~\citenum{Guster2019} on NbSe\s2, we label them as T1 (toward S), ``hexagons'' (toward Nb), T1$'$ (toward gap), and T2$'$ (away from gap). The fact that the experimental d$I$/d$V$ maps largely show a single pronounced peak per $3 \times 3$ cell and are mainly sensitive to the S atoms suggests that the T1 structure is observed in experiment. Thus, we focus our discussion in the main text on the T1 structure (all other structures are described in the SI). To facilitate the comparison, a smearing $\sigma=5$ mRy is used unless otherwise noted, since all four structures are stable at this smearing.

As we are trying to better understand the low-energy d$I$/d$V$ spectra shown in Figure~\ref{fig_sts}, we first consider the calculated electronic structure. Figure~\ref{fig:theory}b--d shows the band structure and electronic DOS of the T1 CDW structure. There is a significant reconstruction compared to the high-temperature undistorted structure with several partial gaps opening mainly in the vicinity of the K~point. Instead of a sharp gap directly at the Fermi level, there is a rather constant depression of the DOS in an interval of about 150\,meV around the Fermi level. Inside this depression region, there are small peaks (Van Hove singularities) whose position is characteristic of the individual CDW structure (see Figures~S9, S10, and S11 in the SI for the other three structures) and the displacement. However, these peaks do not fit the experiment energetically and they are not symmetric around the Fermi level. This suggests that the experimentally observed d$I$/d$V$ is not purely electronic in nature.

One possibility is that inelastic phonon excitations could be responsible for the specific signatures in the STS. Thus, we continue with an analysis of the phononic excitations present in the CDW state.
Once the system has undergone a CDW transition, signatures of the competing CDW structures remain visible in the phonon spectrum. Figure~\ref{fig:theory}e,\,f shows the phonon dispersion in the T1 CDW phase. Highlighted in magenta are phonons corresponding to any of the four displacement patterns in Figure~\ref{fig:theory}a, which are longitudinal-acoustic modes corresponding to the experimentally observed $3\times 3$ periodicity. Of these, the mode with the highest energy is the amplitude (or Higgs) mode, where the atoms oscillate toward their undistorted position and back. The other modes are phase (or Goldstone) modes, corresponding to oscillations toward any of the other CDW patterns. The phase modes have a small but non-zero energy in a dynamically stable commensurate CDW.
The precise energy of these modes depends on the cold smearing parameter $\sigma$. At $\sigma_c=14.7$ mRy, the undistorted structure is on the edge of being stable, and all highlighted phonon modes have precisely zero energy. For $\sigma<\sigma_c$, as in Figure~\ref{fig:theory}e,\,f, the system is in a stable CDW phase. Lowering smearing further, the displacement compared to the symmetric state increases, with a corresponding increase of the phonon energies, \emph{i.e.}, the magenta block in Figure~\ref{fig:theory}e moves upwards for smaller $\sigma$. Figure~\ref{fig:theory}e,\,f and the corresponding panels in the SI have a smearing just below the point where the given structure becomes stable, here $\sigma=13$ mRy, so that the magenta and black modes are clearly separated at $\vec q = \Gamma$. From the point of view of the undistorted structure, the finite energy of the phase modes in the distorted phase is a non-linear phonon-phonon coupling effect. Importantly, if we are sufficiently deep inside the CDW phase, the amplitude and phase mode energies lie robustly within the range of the experimentally observed trough valley (amplitude mode) and the smaller inner-valley peaks (phase modes). This energetic match is generic in the sense that it also applies to the other CDW structures considered in the SI.

To assess the impact of the phonons on the STS, we need to know not only their frequency but also how they couple to the electrons. This is quantified by the Eliashberg function $\alpha^2 F(\omega)$ shown in Figure~\ref{fig:theory}h. The electron-phonon coupling appears squared in the Eliashberg spectral function, since the electron needs to emit and absorb a phonon. The Eliashberg function has a clear onset at the energy corresponding to the lowest phase mode. This shows that the modes corresponding to the longitudinal-acoustic modes at $\vec q = 2/3\,\overline{\Gamma \textrm M}$ in the undistorted state still dominate the coupling in presence of the CDW, due to their large electron-phonon matrix elements~\cite{Lian2023}. On the other hand, the phonon DOS itself has contributions all the way down to zero frequency, coming from the acoustic branches close to $\vec q = \Gamma$, but these are weakly coupled to the electrons, as usual.
The Eliashberg function is qualitatively similar in all four CDW structures (see SI for details) with the onset and peak energies matching the STS spectral features qualitatively. Only the precise quantitative energies of the onset differ between the four structures. The absorption and emission of phonons naturally lead to symmetric structures around the Fermi level, therefore offering an explanation for the main experimental observations.

In summary, we present a comprehensive characterization of the CDW in quasi-freestanding H-\nbs~monolayers grown \emph{in situ} on Gr/Ir(111) by low-temperature STM and STS and by DFT and DFPT calculations. We investigated the electronic footprints and temperature dependence of the $3\times3$ modulation pattern and unambiguously link the modulation to a CDW phenomenon. In high-resolution d$I$/d$V$ spectra, we found a gap with additional features inside. We demonstrated that the gap and features are intertwined with the CDW, given by the new bias locations after doping. The most likely explanation of these low-energy features is not purely electronic, but involves combined electron-phonon quasiparticles where the phase and amplitude phonon modes of the CDW couple to the remaining electronic states at the Fermi level. Our finding of an unconventional CDW gap in monolayer \nbs~provides an alternative perspective on gap opening mechanisms in CDW systems, revealing the role of dynamic effects and lattice fluctuations. These insights underscore the significance of incorporating dynamic lattice effects to accurately interpret the low-energy electronic spectra in CDW or generically ordered systems.

\begin{acknowledgement}
We acknowledge funding from Deutsche Forschungsgemeinschaft (DFG) through CRC~1238 (project number 277146847, subprojects A01 and B06) and EXC~2077 (project number 390741603, University Allowance, University of Bremen). JB gratefully acknowledges the support received from the ``U Bremen Excellence Chair Program'', especially from Lucio Colombi Ciacchi and Nicola Marzari, as well as fruitful discussion with Bogdan Guster. EvL acknowledges support from the Swedish Research Council (Vetenskapsr\aa det, VR) under grant 2022-03090. JF acknowledges financial support from the DFG SPP~2137 (Project FI 2624/1-1).
\end{acknowledgement}

\begin{suppinfo}
Experimental methods (details on the sample growth and STM/STS measurements), low-energy electron diffraction, local DOS (d$I$/d$V$ spectra and band structure calculation) of monolayer \nbs, band structure of \nbs~near $\Gamma$ from quasi-particle scattering, details on FFT filtering of the d$I$/d$V$ maps, doping of Gr underlayer effect on \nbs~CDW, real space visualization of inelastic d$I$/d$V$ features, magnetic field dependence of the inelastic excitations, details on computational methods, generalized free energy, and computational characterization of all four possible CDW patterns.
\end{suppinfo}

\bibliography{ms}

\providecommand{\latin}[1]{#1}
\makeatletter
\providecommand{\doi}
  {\begingroup\let\do\@makeother\dospecials
  \catcode`\{=1 \catcode`\}=2 \doi@aux}
\providecommand{\doi@aux}[1]{\endgroup\texttt{#1}}
\makeatother
\providecommand*\mcitethebibliography{\thebibliography}
\csname @ifundefined\endcsname{endmcitethebibliography}
  {\let\endmcitethebibliography\endthebibliography}{}
\begin{mcitethebibliography}{42}
\providecommand*\natexlab[1]{#1}
\providecommand*\mciteSetBstSublistMode[1]{}
\providecommand*\mciteSetBstMaxWidthForm[2]{}
\providecommand*\mciteBstWouldAddEndPuncttrue
  {\def\EndOfBibitem{\unskip.}}
\providecommand*\mciteBstWouldAddEndPunctfalse
  {\let\EndOfBibitem\relax}
\providecommand*\mciteSetBstMidEndSepPunct[3]{}
\providecommand*\mciteSetBstSublistLabelBeginEnd[3]{}
\providecommand*\EndOfBibitem{}
\mciteSetBstSublistMode{f}
\mciteSetBstMaxWidthForm{subitem}{(\alph{mcitesubitemcount})}
\mciteSetBstSublistLabelBeginEnd
  {\mcitemaxwidthsubitemform\space}
  {\relax}
  {\relax}

\bibitem[Qiu \latin{et~al.}(2021)Qiu, Gong, Wang, Zhang, Yang, Wang, and
  Xiong]{qiu_SC_2021}
Qiu,~D.; Gong,~C.; Wang,~S.; Zhang,~M.; Yang,~C.; Wang,~X.; Xiong,~J. Recent
  {Advances} in {2D} {Superconductors}. \emph{Adv. Mater.} \textbf{2021},
  \emph{33}, 2006124\relax
\mciteBstWouldAddEndPuncttrue
\mciteSetBstMidEndSepPunct{\mcitedefaultmidpunct}
{\mcitedefaultendpunct}{\mcitedefaultseppunct}\relax
\EndOfBibitem
\bibitem[Mak \latin{et~al.}(2019)Mak, Shan, and Ralph]{mak_magnetism_2019}
Mak,~K.~F.; Shan,~J.; Ralph,~D.~C. Probing and controlling magnetic states in
  {2D} layered magnetic materials. \emph{Nat. Rev. Phys.} \textbf{2019},
  \emph{1}, 646--661\relax
\mciteBstWouldAddEndPuncttrue
\mciteSetBstMidEndSepPunct{\mcitedefaultmidpunct}
{\mcitedefaultendpunct}{\mcitedefaultseppunct}\relax
\EndOfBibitem
\bibitem[Rossnagel(2011)]{Rossnagel11}
Rossnagel,~K. On the origin of charge-density waves in select layered
  transition-metal dichalcogenides. \emph{J. Phys.: Condens. Matter}
  \textbf{2011}, \emph{23}, 213001\relax
\mciteBstWouldAddEndPuncttrue
\mciteSetBstMidEndSepPunct{\mcitedefaultmidpunct}
{\mcitedefaultendpunct}{\mcitedefaultseppunct}\relax
\EndOfBibitem
\bibitem[Slater(1951)]{slater_magnetic_1951}
Slater,~J.~C. Magnetic {Effects} and the {Hartree}-{Fock} {Equation}.
  \emph{Phys. Rev.} \textbf{1951}, \emph{82}, 538--541\relax
\mciteBstWouldAddEndPuncttrue
\mciteSetBstMidEndSepPunct{\mcitedefaultmidpunct}
{\mcitedefaultendpunct}{\mcitedefaultseppunct}\relax
\EndOfBibitem
\bibitem[Peierls(1955)]{Peierls55}
Peierls,~R.~E. \emph{Quantum theory of solids}; Clarendon Press: Oxford,
  1955\relax
\mciteBstWouldAddEndPuncttrue
\mciteSetBstMidEndSepPunct{\mcitedefaultmidpunct}
{\mcitedefaultendpunct}{\mcitedefaultseppunct}\relax
\EndOfBibitem
\bibitem[Bardeen \latin{et~al.}(1957)Bardeen, Cooper, and
  Schrieffer]{bardeen_microscopic_1957}
Bardeen,~J.; Cooper,~L.~N.; Schrieffer,~J.~R. Microscopic {Theory} of
  {Superconductivity}. \emph{Phys. Rev.} \textbf{1957}, \emph{106},
  162--164\relax
\mciteBstWouldAddEndPuncttrue
\mciteSetBstMidEndSepPunct{\mcitedefaultmidpunct}
{\mcitedefaultendpunct}{\mcitedefaultseppunct}\relax
\EndOfBibitem
\bibitem[Bardeen \latin{et~al.}(1957)Bardeen, Cooper, and
  Schrieffer]{BCS_Long_1957}
Bardeen,~J.; Cooper,~L.~N.; Schrieffer,~J.~R. Theory of Superconductivity.
  \emph{Phys. Rev.} \textbf{1957}, \emph{108}, 1175--1204\relax
\mciteBstWouldAddEndPuncttrue
\mciteSetBstMidEndSepPunct{\mcitedefaultmidpunct}
{\mcitedefaultendpunct}{\mcitedefaultseppunct}\relax
\EndOfBibitem
\bibitem[Chen \latin{et~al.}(2005)Chen, Stajic, Tan, and Levin]{Chen05}
Chen,~Q.; Stajic,~J.; Tan,~S.; Levin,~K. BCS–BEC crossover: From high
  temperature superconductors to ultracold superfluids. \emph{Phys. Rep.}
  \textbf{2005}, \emph{412}, 1--88\relax
\mciteBstWouldAddEndPuncttrue
\mciteSetBstMidEndSepPunct{\mcitedefaultmidpunct}
{\mcitedefaultendpunct}{\mcitedefaultseppunct}\relax
\EndOfBibitem
\bibitem[Sch\"afer \latin{et~al.}(2015)Sch\"afer, Geles, Rost, Rohringer,
  Arrigoni, Held, Bl\"umer, Aichhorn, and Toschi]{Schafer15}
Sch\"afer,~T.; Geles,~F.; Rost,~D.; Rohringer,~G.; Arrigoni,~E.; Held,~K.;
  Bl\"umer,~N.; Aichhorn,~M.; Toschi,~A. Fate of the false Mott-Hubbard
  transition in two dimensions. \emph{Phys. Rev. B} \textbf{2015}, \emph{91},
  125109\relax
\mciteBstWouldAddEndPuncttrue
\mciteSetBstMidEndSepPunct{\mcitedefaultmidpunct}
{\mcitedefaultendpunct}{\mcitedefaultseppunct}\relax
\EndOfBibitem
\bibitem[Demsar \latin{et~al.}(1999)Demsar, Biljakovi\ifmmode~\acute{c}\else
  \'{c}\fi{}, and Mihailovic]{Demsar99}
Demsar,~J.; Biljakovi\ifmmode~\acute{c}\else \'{c}\fi{},~K.; Mihailovic,~D.
  Single Particle and Collective Excitations in the One-Dimensional Charge
  Density Wave Solid ${\mathrm{K}}_{0.3}{\mathrm{MoO}}_{3}$ Probed in Real Time
  by Femtosecond Spectroscopy. \emph{Phys. Rev. Lett.} \textbf{1999},
  \emph{83}, 800--803\relax
\mciteBstWouldAddEndPuncttrue
\mciteSetBstMidEndSepPunct{\mcitedefaultmidpunct}
{\mcitedefaultendpunct}{\mcitedefaultseppunct}\relax
\EndOfBibitem
\bibitem[Petersen \latin{et~al.}(2011)Petersen, Kaiser, Dean, Simoncig, Liu,
  Cavalieri, Cacho, Turcu, Springate, Frassetto, Poletto, Dhesi, Berger, and
  Cavalleri]{Petersen11}
Petersen,~J.~C.; Kaiser,~S.; Dean,~N.; Simoncig,~A.; Liu,~H.~Y.;
  Cavalieri,~A.~L.; Cacho,~C.; Turcu,~I. C.~E.; Springate,~E.; Frassetto,~F.;
  Poletto,~L.; Dhesi,~S.~S.; Berger,~H.; Cavalleri,~A. Clocking the Melting
  Transition of Charge and Lattice Order in
  $1T\mathrm{\text{\ensuremath{-}}}{\mathrm{TaS}}_{2}$ with Ultrafast
  Extreme-Ultraviolet Angle-Resolved Photoemission Spectroscopy. \emph{Phys.
  Rev. Lett.} \textbf{2011}, \emph{107}, 177402\relax
\mciteBstWouldAddEndPuncttrue
\mciteSetBstMidEndSepPunct{\mcitedefaultmidpunct}
{\mcitedefaultendpunct}{\mcitedefaultseppunct}\relax
\EndOfBibitem
\bibitem[Liu \latin{et~al.}(2013)Liu, Gierz, Petersen, Kaiser, Simoncig,
  Cavalieri, Cacho, Turcu, Springate, Frassetto, Poletto, Dhesi, Xu, Cuk,
  Merlin, and Cavalleri]{Liu13}
Liu,~H.~Y. \latin{et~al.}  Possible observation of parametrically amplified
  coherent phasons in K${}_{0.3}$MoO${}_{3}$ using time-resolved
  extreme-ultraviolet angle-resolved photoemission spectroscopy. \emph{Phys.
  Rev. B} \textbf{2013}, \emph{88}, 045104\relax
\mciteBstWouldAddEndPuncttrue
\mciteSetBstMidEndSepPunct{\mcitedefaultmidpunct}
{\mcitedefaultendpunct}{\mcitedefaultseppunct}\relax
\EndOfBibitem
\bibitem[Johannes and Mazin(2008)Johannes, and Mazin]{Johannes08}
Johannes,~M.~D.; Mazin,~I.~I. Fermi surface nesting and the origin of charge
  density waves in metals. \emph{Phys. Rev. B} \textbf{2008}, \emph{77},
  165135\relax
\mciteBstWouldAddEndPuncttrue
\mciteSetBstMidEndSepPunct{\mcitedefaultmidpunct}
{\mcitedefaultendpunct}{\mcitedefaultseppunct}\relax
\EndOfBibitem
\bibitem[Weber \latin{et~al.}(2011)Weber, Rosenkranz, Castellan, Osborn, Hott,
  Heid, Bohnen, Egami, Said, and Reznik]{Weber11}
Weber,~F.; Rosenkranz,~S.; Castellan,~J.-P.; Osborn,~R.; Hott,~R.; Heid,~R.;
  Bohnen,~K.-P.; Egami,~T.; Said,~A.~H.; Reznik,~D. Extended Phonon Collapse
  and the Origin of the Charge-Density Wave in
  $2H\mathrm{\text{\ensuremath{-}}}{\mathrm{NbSe}}_{2}$. \emph{Phys. Rev.
  Lett.} \textbf{2011}, \emph{107}, 107403\relax
\mciteBstWouldAddEndPuncttrue
\mciteSetBstMidEndSepPunct{\mcitedefaultmidpunct}
{\mcitedefaultendpunct}{\mcitedefaultseppunct}\relax
\EndOfBibitem
\bibitem[Berges \latin{et~al.}(2020)Berges, van Loon, Schobert, R\"osner, and
  Wehling]{Berges20}
Berges,~J.; van Loon,~E. G. C.~P.; Schobert,~A.; R\"osner,~M.; Wehling,~T.~O.
  Ab initio phonon self-energies and fluctuation diagnostics of phonon
  anomalies: Lattice instabilities from Dirac pseudospin physics in transition
  metal dichalcogenides. \emph{Phys. Rev. B} \textbf{2020}, \emph{101},
  155107\relax
\mciteBstWouldAddEndPuncttrue
\mciteSetBstMidEndSepPunct{\mcitedefaultmidpunct}
{\mcitedefaultendpunct}{\mcitedefaultseppunct}\relax
\EndOfBibitem
\bibitem[van Efferen \latin{et~al.}(2021)van Efferen, Berges, Hall, van Loon,
  Kraus, Schobert, Wekking, Huttmann, Plaar, Rothenbach, Ollefs, Arruda,
  Brookes, Sch{\"o}nhoff, Kummer, Wende, Wehling, and Michely]{vanEfferen21}
van Efferen,~C. \latin{et~al.}  A full gap above the Fermi level: the charge
  density wave of monolayer VS$_2$. \emph{Nat. Commun.} \textbf{2021},
  \emph{12}, 6837\relax
\mciteBstWouldAddEndPuncttrue
\mciteSetBstMidEndSepPunct{\mcitedefaultmidpunct}
{\mcitedefaultendpunct}{\mcitedefaultseppunct}\relax
\EndOfBibitem
\bibitem[Hofmann \latin{et~al.}(2019)Hofmann, Ugeda, Tamt\"ogl, Ruckhofer,
  Ernst, Benedek, Mart\'{\i}nez-Galera, Str\'o\ifmmode~\dot{z}\else
  \.{z}\fi{}ecka, G\'omez-Rodr\'{\i}guez, Rienks, Jensen, Pascual, and
  Wells]{Hofmann19}
Hofmann,~P.; Ugeda,~M.~M.; Tamt\"ogl,~A.; Ruckhofer,~A.; Ernst,~W.~E.;
  Benedek,~G.; Mart\'{\i}nez-Galera,~A.~J.; Str\'o\ifmmode~\dot{z}\else
  \.{z}\fi{}ecka,~A.; G\'omez-Rodr\'{\i}guez,~J.~M.; Rienks,~E.; Jensen,~M.~F.;
  Pascual,~J.~I.; Wells,~J.~W. Strong-coupling charge density wave in a
  one-dimensional topological metal. \emph{Phys. Rev. B} \textbf{2019},
  \emph{99}, 035438\relax
\mciteBstWouldAddEndPuncttrue
\mciteSetBstMidEndSepPunct{\mcitedefaultmidpunct}
{\mcitedefaultendpunct}{\mcitedefaultseppunct}\relax
\EndOfBibitem
\bibitem[Fisher and Sienko(1980)Fisher, and Sienko]{Fisher80}
Fisher,~W.~G.; Sienko,~M.~J. Stoichiometry, structure, and physical properties
  of niobium disulfide. \emph{Inorg. Chem.} \textbf{1980}, \emph{19},
  39--43\relax
\mciteBstWouldAddEndPuncttrue
\mciteSetBstMidEndSepPunct{\mcitedefaultmidpunct}
{\mcitedefaultendpunct}{\mcitedefaultseppunct}\relax
\EndOfBibitem
\bibitem[Lin \latin{et~al.}(2018)Lin, Huang, Zhao, Lian, Duan, Chen, and
  Ji]{Lin18}
Lin,~H.; Huang,~W.; Zhao,~K.; Lian,~C.; Duan,~W.; Chen,~X.; Ji,~S.-H. Growth of
  atomically thick transition metal sulfide films on graphene/6H-SiC(0001) by
  molecular beam epitaxy. \emph{Nano Res.} \textbf{2018}, \emph{11},
  4722--4727\relax
\mciteBstWouldAddEndPuncttrue
\mciteSetBstMidEndSepPunct{\mcitedefaultmidpunct}
{\mcitedefaultendpunct}{\mcitedefaultseppunct}\relax
\EndOfBibitem
\bibitem[Stan \latin{et~al.}(2019)Stan, Mahatha, Bianchi, Sanders, Curcio,
  Hofmann, and Miwa]{Stan19}
Stan,~R.-M.; Mahatha,~S.~K.; Bianchi,~M.; Sanders,~C.~E.; Curcio,~D.;
  Hofmann,~P.; Miwa,~J.~A. Epitaxial single-layer ${\mathrm{NbS}}_{2}$ on
  Au(111): Synthesis, structure, and electronic properties. \emph{Phys. Rev.
  Mater.} \textbf{2019}, \emph{3}, 044003\relax
\mciteBstWouldAddEndPuncttrue
\mciteSetBstMidEndSepPunct{\mcitedefaultmidpunct}
{\mcitedefaultendpunct}{\mcitedefaultseppunct}\relax
\EndOfBibitem
\bibitem[Ugeda \latin{et~al.}(2016)Ugeda, Bradley, Zhang, Onishi, Chen, Ruan,
  Ojeda-Aristizabal, Ryu, Edmonds, Tsai, Riss, Mo, Lee, Zettl, Hussain, Shen,
  and Crommie]{Ugeda16}
Ugeda,~M.~M. \latin{et~al.}  Characterization of collective ground states in
  single-layer NbSe$_2$. \emph{Nat. Phys.} \textbf{2016}, \emph{12},
  92--97\relax
\mciteBstWouldAddEndPuncttrue
\mciteSetBstMidEndSepPunct{\mcitedefaultmidpunct}
{\mcitedefaultendpunct}{\mcitedefaultseppunct}\relax
\EndOfBibitem
\bibitem[Naito and Tanaka(1982)Naito, and Tanaka]{Naito82}
Naito,~M.; Tanaka,~S. Electrical Transport Properties in 2H-NbS$_2$, -NbSe$_2$,
  -TaS$_2$ and -TaSe$_2$. \emph{J. Phys. Soc. Jpn.} \textbf{1982}, \emph{51},
  219--227\relax
\mciteBstWouldAddEndPuncttrue
\mciteSetBstMidEndSepPunct{\mcitedefaultmidpunct}
{\mcitedefaultendpunct}{\mcitedefaultseppunct}\relax
\EndOfBibitem
\bibitem[Guillam\'on \latin{et~al.}(2008)Guillam\'on, Suderow, Vieira, Cario,
  Diener, and Rodi\`ere]{Guillamon08}
Guillam\'on,~I.; Suderow,~H.; Vieira,~S.; Cario,~L.; Diener,~P.; Rodi\`ere,~P.
  Superconducting Density of States and Vortex Cores of
  2H-${\mathrm{NbS}}_{2}$. \emph{Phys. Rev. Lett.} \textbf{2008}, \emph{101},
  166407\relax
\mciteBstWouldAddEndPuncttrue
\mciteSetBstMidEndSepPunct{\mcitedefaultmidpunct}
{\mcitedefaultendpunct}{\mcitedefaultseppunct}\relax
\EndOfBibitem
\bibitem[Heil \latin{et~al.}(2017)Heil, Ponc\'e, Lambert, Schlipf, Margine, and
  Giustino]{Heil17}
Heil,~C.; Ponc\'e,~S.; Lambert,~H.; Schlipf,~M.; Margine,~E.~R.; Giustino,~F.
  Origin of Superconductivity and Latent Charge Density Wave in
  ${\mathrm{NbS}}_{2}$. \emph{Phys. Rev. Lett.} \textbf{2017}, \emph{119},
  087003\relax
\mciteBstWouldAddEndPuncttrue
\mciteSetBstMidEndSepPunct{\mcitedefaultmidpunct}
{\mcitedefaultendpunct}{\mcitedefaultseppunct}\relax
\EndOfBibitem
\bibitem[Bianco \latin{et~al.}(2019)Bianco, Errea, Monacelli, Calandra, and
  Mauri]{Bianco19}
Bianco,~R.; Errea,~I.; Monacelli,~L.; Calandra,~M.; Mauri,~F. Quantum
  Enhancement of Charge Density Wave in NbS$_2$ in the Two-Dimensional Limit.
  \emph{Nano Letters} \textbf{2019}, \emph{19}, 3098--3103\relax
\mciteBstWouldAddEndPuncttrue
\mciteSetBstMidEndSepPunct{\mcitedefaultmidpunct}
{\mcitedefaultendpunct}{\mcitedefaultseppunct}\relax
\EndOfBibitem
\bibitem[van Efferen \latin{et~al.}(2022)van Efferen, Murray, Fischer, Busse,
  Komsa, Michely, and Jolie]{vanEfferen22}
van Efferen,~C.; Murray,~C.; Fischer,~J.; Busse,~C.; Komsa,~H.-P.; Michely,~T.;
  Jolie,~W. Metal-insulator transition in monolayer MoS$_2$ via contactless
  chemical doping. \emph{2D Mater.} \textbf{2022}, \emph{9}, 025026\relax
\mciteBstWouldAddEndPuncttrue
\mciteSetBstMidEndSepPunct{\mcitedefaultmidpunct}
{\mcitedefaultendpunct}{\mcitedefaultseppunct}\relax
\EndOfBibitem
\bibitem[Ryu \latin{et~al.}(2018)Ryu, Chen, Kim, Tsai, Tang, Jiang, Liou, Kahn,
  Jia, Omrani, Shim, Hussain, Shen, Kim, Min, Hwang, Crommie, and Mo]{Ryu17}
Ryu,~H. \latin{et~al.}  Persistent Charge-Density-Wave Order in Single-Layer
  TaSe$_2$. \emph{Nano Lett.} \textbf{2018}, \emph{18}, 689--694, PMID:
  29300484\relax
\mciteBstWouldAddEndPuncttrue
\mciteSetBstMidEndSepPunct{\mcitedefaultmidpunct}
{\mcitedefaultendpunct}{\mcitedefaultseppunct}\relax
\EndOfBibitem
\bibitem[Wan \latin{et~al.}(2022)Wan, Dreher, Muñoz-Segovia, Harsh, Guo,
  Martínez-Galera, Guinea, de~Juan, and Ugeda]{Wan22}
Wan,~W.; Dreher,~P.; Muñoz-Segovia,~D.; Harsh,~R.; Guo,~H.;
  Martínez-Galera,~A.~J.; Guinea,~F.; de~Juan,~F.; Ugeda,~M.~M. Observation of
  Superconducting Collective Modes from Competing Pairing Instabilities in
  Single-Layer NbSe$_2$. \emph{Adv. Mater.} \textbf{2022}, \emph{34},
  2206078\relax
\mciteBstWouldAddEndPuncttrue
\mciteSetBstMidEndSepPunct{\mcitedefaultmidpunct}
{\mcitedefaultendpunct}{\mcitedefaultseppunct}\relax
\EndOfBibitem
\bibitem[Zhang \latin{et~al.}(2008)Zhang, Brar, Wang, Girit, Yayon, Panlasigui,
  Zettl, and Crommie]{Zhang08}
Zhang,~Y.; Brar,~V.~W.; Wang,~F.; Girit,~C.; Yayon,~Y.; Panlasigui,~M.;
  Zettl,~A.; Crommie,~M.~F. Giant phonon-induced conductance in scanning
  tunnelling spectroscopy of gate-tunable graphene. \emph{Nat. Phys.}
  \textbf{2008}, \emph{4}, 627--630\relax
\mciteBstWouldAddEndPuncttrue
\mciteSetBstMidEndSepPunct{\mcitedefaultmidpunct}
{\mcitedefaultendpunct}{\mcitedefaultseppunct}\relax
\EndOfBibitem
\bibitem[Gawronski \latin{et~al.}(2008)Gawronski, Mehlhorn, and
  Morgenstern]{Gawronski08}
Gawronski,~H.; Mehlhorn,~M.; Morgenstern,~K. Imaging Phonon Excitation with
  Atomic Resolution. \emph{Science} \textbf{2008}, \emph{319}, 930--933\relax
\mciteBstWouldAddEndPuncttrue
\mciteSetBstMidEndSepPunct{\mcitedefaultmidpunct}
{\mcitedefaultendpunct}{\mcitedefaultseppunct}\relax
\EndOfBibitem
\bibitem[Nishio \latin{et~al.}(1994)Nishio, Shirai, Suzuki, and
  Motizuki]{Nishio94}
Nishio,~Y.; Shirai,~M.; Suzuki,~N.; Motizuki,~K. Role of Electron-Lattice
  Interaction in Layered Transition Metal Dichalcogenide $\mathrm{2H-NbS_2}$.
  I. Phonon Anomaly and Superconductivity. \emph{J. Phys. Soc. Jpn.}
  \textbf{1994}, \emph{63}, 156--167\relax
\mciteBstWouldAddEndPuncttrue
\mciteSetBstMidEndSepPunct{\mcitedefaultmidpunct}
{\mcitedefaultendpunct}{\mcitedefaultseppunct}\relax
\EndOfBibitem
\bibitem[Wen \latin{et~al.}(2020)Wen, Xie, Wu, Shen, Kong, Lian, Li, Xing, and
  Yan]{Wen20}
Wen,~C.; Xie,~Y.; Wu,~Y.; Shen,~S.; Kong,~P.; Lian,~H.; Li,~J.; Xing,~H.;
  Yan,~S. Impurity-pinned incommensurate charge density wave and local phonon
  excitations in $2H\ensuremath{-}{\mathrm{NbS}}_{2}$. \emph{Phys. Rev. B}
  \textbf{2020}, \emph{101}, 241404\relax
\mciteBstWouldAddEndPuncttrue
\mciteSetBstMidEndSepPunct{\mcitedefaultmidpunct}
{\mcitedefaultendpunct}{\mcitedefaultseppunct}\relax
\EndOfBibitem
\bibitem[Hou \latin{et~al.}(2020)Hou, Zhang, Tu, Gu, Zhang, Gong, Tu, Wang, Lv,
  Weng, Ren, Chen, Zhu, Hao, and Shan]{Hou20}
Hou,~X.-Y.; Zhang,~F.; Tu,~X.-H.; Gu,~Y.-D.; Zhang,~M.-D.; Gong,~J.; Tu,~Y.-B.;
  Wang,~B.-T.; Lv,~W.-G.; Weng,~H.-M.; Ren,~Z.-A.; Chen,~G.-F.; Zhu,~X.-D.;
  Hao,~N.; Shan,~L. Inelastic Electron Tunneling in
  $2\mathrm{H}\text{\ensuremath{-}}{\mathrm{Ta}}_{x}{\mathrm{Nb}}_{1\ensuremath{-}x}{\mathrm{Se}}_{2}$
  Evidenced by Scanning Tunneling Spectroscopy. \emph{Phys. Rev. Lett.}
  \textbf{2020}, \emph{124}, 106403\relax
\mciteBstWouldAddEndPuncttrue
\mciteSetBstMidEndSepPunct{\mcitedefaultmidpunct}
{\mcitedefaultendpunct}{\mcitedefaultseppunct}\relax
\EndOfBibitem
\bibitem[Xu \latin{et~al.}(2014)Xu, Liu, and Guo]{Xu14}
Xu,~Y.; Liu,~X.; Guo,~W. Tensile strain induced switching of magnetic states in
  NbSe$_2$ and NbS$_2$ single layers. \emph{Nanoscale} \textbf{2014}, \emph{6},
  12929--12933\relax
\mciteBstWouldAddEndPuncttrue
\mciteSetBstMidEndSepPunct{\mcitedefaultmidpunct}
{\mcitedefaultendpunct}{\mcitedefaultseppunct}\relax
\EndOfBibitem
\bibitem[G\"uller \latin{et~al.}(2016)G\"uller, Vildosola, and Llois]{Guller16}
G\"uller,~F.; Vildosola,~V.~L.; Llois,~A.~M. Spin density wave instabilities in
  the ${\mathrm{NbS}}_{2}$ monolayer. \emph{Phys. Rev. B} \textbf{2016},
  \emph{93}, 094434\relax
\mciteBstWouldAddEndPuncttrue
\mciteSetBstMidEndSepPunct{\mcitedefaultmidpunct}
{\mcitedefaultendpunct}{\mcitedefaultseppunct}\relax
\EndOfBibitem
\bibitem[van Loon \latin{et~al.}(2018)van Loon, R{\"o}sner, Sch{\"o}nhoff,
  Katsnelson, and Wehling]{vanloon18}
van Loon,~E. G. C.~P.; R{\"o}sner,~M.; Sch{\"o}nhoff,~G.; Katsnelson,~M.~I.;
  Wehling,~T.~O. Competing Coulomb and electron--phonon interactions in
  NbS$_2$. \emph{npj Quantum Mater.} \textbf{2018}, \emph{3}, 32\relax
\mciteBstWouldAddEndPuncttrue
\mciteSetBstMidEndSepPunct{\mcitedefaultmidpunct}
{\mcitedefaultendpunct}{\mcitedefaultseppunct}\relax
\EndOfBibitem
\bibitem[Schobert \latin{et~al.}(2023)Schobert, Berges, van Loon, Sentef,
  Brener, Rossi, and Wehling]{Schobert2023}
Schobert,~A.; Berges,~J.; van Loon,~E. G. C.~P.; Sentef,~M.~A.; Brener,~S.;
  Rossi,~M.; Wehling,~T.~O. Ab initio electron-lattice downfolding: Potential
  energy landscapes, anharmonicity, and molecular dynamics in charge density
  wave materials. https://arxiv.org/abs/2303.07261, 2023\relax
\mciteBstWouldAddEndPuncttrue
\mciteSetBstMidEndSepPunct{\mcitedefaultmidpunct}
{\mcitedefaultendpunct}{\mcitedefaultseppunct}\relax
\EndOfBibitem
\bibitem[Marzari \latin{et~al.}(1999)Marzari, Vanderbilt, De~Vita, and
  Payne]{Marzari1999}
Marzari,~N.; Vanderbilt,~D.; De~Vita,~A.; Payne,~M.~C. Thermal Contraction and
  Disordering of the {Al}(110) Surface. \emph{Phys. Rev. Lett.} \textbf{1999},
  \emph{82}, 3296\relax
\mciteBstWouldAddEndPuncttrue
\mciteSetBstMidEndSepPunct{\mcitedefaultmidpunct}
{\mcitedefaultendpunct}{\mcitedefaultseppunct}\relax
\EndOfBibitem
\bibitem[Hall \latin{et~al.}(2019)Hall, Ehlen, Berges, van Loon, van Efferen,
  Murray, R\"osner, Li, Senkovskiy, Hell, Rolf, Heider, Asensio, Avila,
  Plucinski, Wehling, Gr\"uneis, and Michely]{Hall19}
Hall,~J. \latin{et~al.}  Environmental Control of Charge Density Wave Order in
  Monolayer 2H-TaS$_2$. \emph{ACS Nano} \textbf{2019}, \emph{13},
  10210--10220\relax
\mciteBstWouldAddEndPuncttrue
\mciteSetBstMidEndSepPunct{\mcitedefaultmidpunct}
{\mcitedefaultendpunct}{\mcitedefaultseppunct}\relax
\EndOfBibitem
\bibitem[Guster \latin{et~al.}(2019)Guster, Rubio-Verd\'u, Robles, Zald\'ivar,
  Dreher, Pruneda, Silva-Guill\'en, Choi, Pascual, Ugeda, Ordej\'on, and
  Canadell]{Guster2019}
Guster,~B.; Rubio-Verd\'u,~C.; Robles,~R.; Zald\'ivar,~J.; Dreher,~P.;
  Pruneda,~M.; Silva-Guill\'en,~J.~A.; Choi,~D.-J.; Pascual,~J.~I.;
  Ugeda,~M.~M.; Ordej\'on,~P.; Canadell,~E. Coexistence of Elastic Modulations
  in the Charge Density Wave State of {2H}-{NbSe$_2$}. \emph{Nano Lett.}
  \textbf{2019}, \emph{19}, 3027\relax
\mciteBstWouldAddEndPuncttrue
\mciteSetBstMidEndSepPunct{\mcitedefaultmidpunct}
{\mcitedefaultendpunct}{\mcitedefaultseppunct}\relax
\EndOfBibitem
\bibitem[Lian(2023)]{Lian2023}
Lian,~C.-S. Interplay of charge ordering and superconductivity in
  two-dimensional {2H} group {V} transition-metal dichalcogenides. \emph{Phys.
  Rev. B} \textbf{2023}, \emph{107}, 045431\relax
\mciteBstWouldAddEndPuncttrue
\mciteSetBstMidEndSepPunct{\mcitedefaultmidpunct}
{\mcitedefaultendpunct}{\mcitedefaultseppunct}\relax
\EndOfBibitem
\end{mcitethebibliography}


\providecommand{\latin}[1]{#1}
\makeatletter
\providecommand{\doi}
  {\begingroup\let\do\@makeother\dospecials
  \catcode`\{=1 \catcode`\}=2 \doi@aux}
\providecommand{\doi@aux}[1]{\endgroup\texttt{#1}}
\makeatother
\providecommand*\mcitethebibliography{\thebibliography}
\csname @ifundefined\endcsname{endmcitethebibliography}
  {\let\endmcitethebibliography\endthebibliography}{}
\begin{mcitethebibliography}{29}
\providecommand*\natexlab[1]{#1}
\providecommand*\mciteSetBstSublistMode[1]{}
\providecommand*\mciteSetBstMaxWidthForm[2]{}
\providecommand*\mciteBstWouldAddEndPuncttrue
  {\def\EndOfBibitem{\unskip.}}
\providecommand*\mciteBstWouldAddEndPunctfalse
  {\let\EndOfBibitem\relax}
\providecommand*\mciteSetBstMidEndSepPunct[3]{}
\providecommand*\mciteSetBstSublistLabelBeginEnd[3]{}
\providecommand*\EndOfBibitem{}
\mciteSetBstSublistMode{f}
\mciteSetBstMaxWidthForm{subitem}{(\alph{mcitesubitemcount})}
\mciteSetBstSublistLabelBeginEnd
  {\mcitemaxwidthsubitemform\space}
  {\relax}
  {\relax}

\bibitem[Hall \latin{et~al.}(2018)Hall, Pieli{\'{c}}, Murray, Jolie, Wekking,
  Busse, Kralj, and Michely]{Hall18}
Hall,~J.; Pieli{\'{c}},~B.; Murray,~C.; Jolie,~W.; Wekking,~T.; Busse,~C.;
  Kralj,~M.; Michely,~T. Molecular beam epitaxy of quasi-freestanding
  transition metal disulphide monolayers on van der Waals substrates: a growth
  study. \emph{2D Mater.} \textbf{2018}, \emph{5}, 025005\relax
\mciteBstWouldAddEndPuncttrue
\mciteSetBstMidEndSepPunct{\mcitedefaultmidpunct}
{\mcitedefaultendpunct}{\mcitedefaultseppunct}\relax
\EndOfBibitem
\bibitem[Kaiser and Jaklevic(1986)Kaiser, and Jaklevic]{Kaiser86}
Kaiser,~W.~J.; Jaklevic,~R.~C. Spectroscopy of electronic states of metals with
  a scanning tunneling microscope. \emph{IBM J. Res. Dev.} \textbf{1986},
  \emph{30}, 411--416\relax
\mciteBstWouldAddEndPuncttrue
\mciteSetBstMidEndSepPunct{\mcitedefaultmidpunct}
{\mcitedefaultendpunct}{\mcitedefaultseppunct}\relax
\EndOfBibitem
\bibitem[M.~P.~Everson and Shen(1991)M.~P.~Everson, and Shen]{Everson91}
M.~P.~Everson,~R. C.~J.,~L. C.~Davis; Shen,~W. Effects of surface features upon
  the Au(111) surface state local density of states studied with scanning
  tunneling spectroscopy. \emph{J. Vac. Sci. Technol. B} \textbf{1991},
  \emph{9}, 891--896\relax
\mciteBstWouldAddEndPuncttrue
\mciteSetBstMidEndSepPunct{\mcitedefaultmidpunct}
{\mcitedefaultendpunct}{\mcitedefaultseppunct}\relax
\EndOfBibitem
\bibitem[Feenstra \latin{et~al.}(1987)Feenstra, Stroscio, and Fein]{Feenstra87}
Feenstra,~R.; Stroscio,~J.~A.; Fein,~A. Tunneling spectroscopy of the Si(111)2
  $\times$ 1 surface. \emph{Surf. Sci.} \textbf{1987}, \emph{181},
  295--306\relax
\mciteBstWouldAddEndPuncttrue
\mciteSetBstMidEndSepPunct{\mcitedefaultmidpunct}
{\mcitedefaultendpunct}{\mcitedefaultseppunct}\relax
\EndOfBibitem
\bibitem[Tersoff and Hamann(1983)Tersoff, and Hamann]{Tersoff83}
Tersoff,~J.; Hamann,~D.~R. Theory and Application for the Scanning Tunneling
  Microscope. \emph{Phys. Rev. Lett.} \textbf{1983}, \emph{50},
  1998--2001\relax
\mciteBstWouldAddEndPuncttrue
\mciteSetBstMidEndSepPunct{\mcitedefaultmidpunct}
{\mcitedefaultendpunct}{\mcitedefaultseppunct}\relax
\EndOfBibitem
\bibitem[Petersen \latin{et~al.}(1998)Petersen, Sprunger, Hofmann,
  L\ae{}gsgaard, Briner, Doering, Rust, Bradshaw, Besenbacher, and
  Plummer]{Petersen98}
Petersen,~L.; Sprunger,~P.~T.; Hofmann,~P.; L\ae{}gsgaard,~E.; Briner,~B.~G.;
  Doering,~M.; Rust,~H.-P.; Bradshaw,~A.~M.; Besenbacher,~F.; Plummer,~E.~W.
  Direct imaging of the two-dimensional Fermi contour: Fourier-transform STM.
  \emph{Phys. Rev. B} \textbf{1998}, \emph{57}, R6858--R6861\relax
\mciteBstWouldAddEndPuncttrue
\mciteSetBstMidEndSepPunct{\mcitedefaultmidpunct}
{\mcitedefaultendpunct}{\mcitedefaultseppunct}\relax
\EndOfBibitem
\bibitem[Crommie \latin{et~al.}(1993)Crommie, Lutz, and Eigler]{Crommie93}
Crommie,~M.~F.; Lutz,~C.~P.; Eigler,~D.~M. Imaging standing waves in a
  two-dimensional electron gas. \emph{Nature} \textbf{1993}, \emph{363},
  524--527\relax
\mciteBstWouldAddEndPuncttrue
\mciteSetBstMidEndSepPunct{\mcitedefaultmidpunct}
{\mcitedefaultendpunct}{\mcitedefaultseppunct}\relax
\EndOfBibitem
\bibitem[Hasegawa and Avouris(1993)Hasegawa, and Avouris]{Hasegawa93}
Hasegawa,~Y.; Avouris,~P. Direct observation of standing wave formation at
  surface steps using scanning tunneling spectroscopy. \emph{Phys. Rev. Lett.}
  \textbf{1993}, \emph{71}, 1071--1074\relax
\mciteBstWouldAddEndPuncttrue
\mciteSetBstMidEndSepPunct{\mcitedefaultmidpunct}
{\mcitedefaultendpunct}{\mcitedefaultseppunct}\relax
\EndOfBibitem
\bibitem[H\"ormandinger(1994)]{Hormandinger94}
H\"ormandinger,~G. Imaging of the Cu(111) surface state in scanning tunneling
  microscopy. \emph{Phys. Rev. B} \textbf{1994}, \emph{49}, 13897--13905\relax
\mciteBstWouldAddEndPuncttrue
\mciteSetBstMidEndSepPunct{\mcitedefaultmidpunct}
{\mcitedefaultendpunct}{\mcitedefaultseppunct}\relax
\EndOfBibitem
\bibitem[Martínez-Galera \latin{et~al.}(2016)Martínez-Galera, Schröder,
  Huttmann, Jolie, Craes, Busse, Caciuc, Atodiresei, Blügel, and
  Michely]{Galera16}
Martínez-Galera,~A.~J.; Schröder,~U.~A.; Huttmann,~F.; Jolie,~W.; Craes,~F.;
  Busse,~C.; Caciuc,~V.; Atodiresei,~N.; Blügel,~S.; Michely,~T. Oxygen orders
  differently under graphene: new superstructures on Ir(111). \emph{Nanoscale}
  \textbf{2016}, \emph{8}, 1932--1943\relax
\mciteBstWouldAddEndPuncttrue
\mciteSetBstMidEndSepPunct{\mcitedefaultmidpunct}
{\mcitedefaultendpunct}{\mcitedefaultseppunct}\relax
\EndOfBibitem
\bibitem[van Efferen \latin{et~al.}(2022)van Efferen, Murray, Fischer, Busse,
  Komsa, Michely, and Jolie]{vanEfferen22}
van Efferen,~C.; Murray,~C.; Fischer,~J.; Busse,~C.; Komsa,~H.-P.; Michely,~T.;
  Jolie,~W. Metal-insulator transition in monolayer MoS$_2$ via contactless
  chemical doping. \emph{2D Mater.} \textbf{2022}, \emph{9}, 025026\relax
\mciteBstWouldAddEndPuncttrue
\mciteSetBstMidEndSepPunct{\mcitedefaultmidpunct}
{\mcitedefaultendpunct}{\mcitedefaultseppunct}\relax
\EndOfBibitem
\bibitem[Marzari \latin{et~al.}(1999)Marzari, Vanderbilt, De~Vita, and
  Payne]{Marzari1999}
Marzari,~N.; Vanderbilt,~D.; De~Vita,~A.; Payne,~M.~C. Thermal Contraction and
  Disordering of the {Al}(110) Surface. \emph{Phys. Rev. Lett.} \textbf{1999},
  \emph{82}, 3296\relax
\mciteBstWouldAddEndPuncttrue
\mciteSetBstMidEndSepPunct{\mcitedefaultmidpunct}
{\mcitedefaultendpunct}{\mcitedefaultseppunct}\relax
\EndOfBibitem
\bibitem[Hall \latin{et~al.}(2019)Hall, Ehlen, Berges, van Loon, van Efferen,
  Murray, R\"osner, Li, Senkovskiy, Hell, Rolf, Heider, Asensio, Avila,
  Plucinski, Wehling, Gr\"uneis, and Michely]{Hall19}
Hall,~J. \latin{et~al.}  Environmental Control of Charge Density Wave Order in
  Monolayer 2H-TaS$_2$. \emph{ACS Nano} \textbf{2019}, \emph{13},
  10210--10220\relax
\mciteBstWouldAddEndPuncttrue
\mciteSetBstMidEndSepPunct{\mcitedefaultmidpunct}
{\mcitedefaultendpunct}{\mcitedefaultseppunct}\relax
\EndOfBibitem
\bibitem[Giannozzi \latin{et~al.}(2009)Giannozzi, \latin{et~al.}
  others]{Giannozzi2009}
Giannozzi,~P.; others \textsc{Quantum ESPRESSO}: A modular and open-source
  software project for quantum simulations of materials. \emph{J. Phys.
  Condens. Matter} \textbf{2009}, \emph{21}, 395502\relax
\mciteBstWouldAddEndPuncttrue
\mciteSetBstMidEndSepPunct{\mcitedefaultmidpunct}
{\mcitedefaultendpunct}{\mcitedefaultseppunct}\relax
\EndOfBibitem
\bibitem[Giannozzi \latin{et~al.}(2017)Giannozzi, \latin{et~al.}
  others]{Giannozzi2017}
Giannozzi,~P.; others Advanced capabilities for materials modelling with
  \textsc{Quantum ESPRESSO}. \emph{J. Phys. Condens. Matter} \textbf{2017},
  \emph{29}, 465901\relax
\mciteBstWouldAddEndPuncttrue
\mciteSetBstMidEndSepPunct{\mcitedefaultmidpunct}
{\mcitedefaultendpunct}{\mcitedefaultseppunct}\relax
\EndOfBibitem
\bibitem[Giannozzi \latin{et~al.}(2020)Giannozzi, Baseggio, Bonf\`a, Brunato,
  Car, Carnimeo, Cavazzoni, de~Gironcoli, Delugas, Ferrari~Ruffino, Ferretti,
  Marzari, Timrov, Urru, and Baroni]{Giannozzi2020}
Giannozzi,~P.; Baseggio,~O.; Bonf\`a,~P.; Brunato,~D.; Car,~R.; Carnimeo,~I.;
  Cavazzoni,~C.; de~Gironcoli,~S.; Delugas,~P.; Ferrari~Ruffino,~F.;
  Ferretti,~A.; Marzari,~N.; Timrov,~I.; Urru,~A.; Baroni,~S. Quantum
  {ESPRESSO} toward the exascale. \emph{J. Chem. Phys.} \textbf{2020},
  \emph{152}, 154105\relax
\mciteBstWouldAddEndPuncttrue
\mciteSetBstMidEndSepPunct{\mcitedefaultmidpunct}
{\mcitedefaultendpunct}{\mcitedefaultseppunct}\relax
\EndOfBibitem
\bibitem[Perdew \latin{et~al.}(1996)Perdew, Burke, and Ernzerhof]{Perdew1996}
Perdew,~J.~P.; Burke,~K.; Ernzerhof,~M. Generalized Gradient Approximation Made
  Simple. \emph{Phys. Rev. Lett.} \textbf{1996}, \emph{77}, 3865\relax
\mciteBstWouldAddEndPuncttrue
\mciteSetBstMidEndSepPunct{\mcitedefaultmidpunct}
{\mcitedefaultendpunct}{\mcitedefaultseppunct}\relax
\EndOfBibitem
\bibitem[Hamann(2013)]{Hamann2013}
Hamann,~D.~R. Optimized norm-conserving {Vanderbilt} pseudopotentials.
  \emph{Phys. Rev. B} \textbf{2013}, \emph{88}, 085117\relax
\mciteBstWouldAddEndPuncttrue
\mciteSetBstMidEndSepPunct{\mcitedefaultmidpunct}
{\mcitedefaultendpunct}{\mcitedefaultseppunct}\relax
\EndOfBibitem
\bibitem[van Setten \latin{et~al.}(2018)van Setten, Giantomassi, Bousquet,
  Verstraete, Hamann, Gonze, and Rignanese]{vanSetten2018}
van Setten,~M.~J.; Giantomassi,~M.; Bousquet,~E.; Verstraete,~M.~J.;
  Hamann,~D.~R.; Gonze,~X.; Rignanese,~G.~M. The {PseudoDojo}: Training and
  grading a 85 element optimized norm-conserving pseudopotential table.
  \emph{Comput. Phys. Commun.} \textbf{2018}, \emph{226}, 39\relax
\mciteBstWouldAddEndPuncttrue
\mciteSetBstMidEndSepPunct{\mcitedefaultmidpunct}
{\mcitedefaultendpunct}{\mcitedefaultseppunct}\relax
\EndOfBibitem
\bibitem[Calandra \latin{et~al.}(2010)Calandra, Profeta, and
  Mauri]{Calandra2010}
Calandra,~M.; Profeta,~G.; Mauri,~F. Adiabatic and nonadiabatic phonon
  dispersion in a {Wannier} function approach. \emph{Phys. Rev. B}
  \textbf{2010}, \emph{82}, 165111\relax
\mciteBstWouldAddEndPuncttrue
\mciteSetBstMidEndSepPunct{\mcitedefaultmidpunct}
{\mcitedefaultendpunct}{\mcitedefaultseppunct}\relax
\EndOfBibitem
\bibitem[Berges \latin{et~al.}(2022)Berges, Girotto, Wehling, Marzari, and
  Ponc\'e]{Berges2022}
Berges,~J.; Girotto,~N.; Wehling,~T.; Marzari,~N.; Ponc\'e,~S. Phonon
  self-energy corrections: To screen, or not to screen.
  https://arxiv.org/abs/2212.11806, 2022\relax
\mciteBstWouldAddEndPuncttrue
\mciteSetBstMidEndSepPunct{\mcitedefaultmidpunct}
{\mcitedefaultendpunct}{\mcitedefaultseppunct}\relax
\EndOfBibitem
\bibitem[Sohier \latin{et~al.}(2017)Sohier, Calandra, and Mauri]{Sohier2017}
Sohier,~T.; Calandra,~M.; Mauri,~F. Density functional perturbation theory for
  gated two-dimensional heterostructures: Theoretical developments and
  application to flexural phonons in graphene. \emph{Phys. Rev. B}
  \textbf{2017}, \emph{96}, 075448\relax
\mciteBstWouldAddEndPuncttrue
\mciteSetBstMidEndSepPunct{\mcitedefaultmidpunct}
{\mcitedefaultendpunct}{\mcitedefaultseppunct}\relax
\EndOfBibitem
\bibitem[Pizzi \latin{et~al.}(2020)Pizzi, \latin{et~al.} others]{Pizzi2020}
Pizzi,~G.; others {Wannier90} as a community code: New features and
  applications. \emph{J. Phys. Condens. Matter} \textbf{2020}, \emph{32},
  165902\relax
\mciteBstWouldAddEndPuncttrue
\mciteSetBstMidEndSepPunct{\mcitedefaultmidpunct}
{\mcitedefaultendpunct}{\mcitedefaultseppunct}\relax
\EndOfBibitem
\bibitem[Giustino \latin{et~al.}(2007)Giustino, Cohen, and Louie]{Giustino2007}
Giustino,~F.; Cohen,~M.~L.; Louie,~S.~G. Electron-phonon interaction using
  {Wannier} functions. \emph{Phys. Rev. B} \textbf{2007}, \emph{76},
  165108\relax
\mciteBstWouldAddEndPuncttrue
\mciteSetBstMidEndSepPunct{\mcitedefaultmidpunct}
{\mcitedefaultendpunct}{\mcitedefaultseppunct}\relax
\EndOfBibitem
\bibitem[Noffsinger \latin{et~al.}(2010)Noffsinger, Giustino, Malone, Park,
  Louie, and Cohen]{Noffsinger2010}
Noffsinger,~J.; Giustino,~F.; Malone,~B.~D.; Park,~C.-H.; Louie,~S.~G.;
  Cohen,~M.~L. {EPW}: A program for calculating the electron--phonon coupling
  using maximally localized {Wannier} functions. \emph{Comput. Phys. Commun.}
  \textbf{2010}, \emph{181}, 2140\relax
\mciteBstWouldAddEndPuncttrue
\mciteSetBstMidEndSepPunct{\mcitedefaultmidpunct}
{\mcitedefaultendpunct}{\mcitedefaultseppunct}\relax
\EndOfBibitem
\bibitem[Ponc\'e \latin{et~al.}(2016)Ponc\'e, Margine, Verdi, and
  Giustino]{Ponce2016}
Ponc\'e,~S.; Margine,~E.; Verdi,~C.; Giustino,~F. {EPW}: Electron--phonon
  coupling, transport and superconducting properties using maximally localized
  {Wannier} functions. \emph{Comput. Phys. Commun.} \textbf{2016}, \emph{209},
  116\relax
\mciteBstWouldAddEndPuncttrue
\mciteSetBstMidEndSepPunct{\mcitedefaultmidpunct}
{\mcitedefaultendpunct}{\mcitedefaultseppunct}\relax
\EndOfBibitem
\bibitem[Guster \latin{et~al.}(2019)Guster, Rubio-Verd\'u, Robles, Zald\'ivar,
  Dreher, Pruneda, Silva-Guill\'en, Choi, Pascual, Ugeda, Ordej\'on, and
  Canadell]{Guster2019}
Guster,~B.; Rubio-Verd\'u,~C.; Robles,~R.; Zald\'ivar,~J.; Dreher,~P.;
  Pruneda,~M.; Silva-Guill\'en,~J.~A.; Choi,~D.-J.; Pascual,~J.~I.;
  Ugeda,~M.~M.; Ordej\'on,~P.; Canadell,~E. Coexistence of Elastic Modulations
  in the Charge Density Wave State of {2H}-{NbSe$_2$}. \emph{Nano Lett.}
  \textbf{2019}, \emph{19}, 3027\relax
\mciteBstWouldAddEndPuncttrue
\mciteSetBstMidEndSepPunct{\mcitedefaultmidpunct}
{\mcitedefaultendpunct}{\mcitedefaultseppunct}\relax
\EndOfBibitem
\bibitem[Lian(2023)]{Lian2023}
Lian,~C.-S. Interplay of charge ordering and superconductivity in
  two-dimensional {2H} group {V} transition-metal dichalcogenides. \emph{Phys.
  Rev. B} \textbf{2023}, \emph{107}, 045431\relax
\mciteBstWouldAddEndPuncttrue
\mciteSetBstMidEndSepPunct{\mcitedefaultmidpunct}
{\mcitedefaultendpunct}{\mcitedefaultseppunct}\relax
\EndOfBibitem
\end{mcitethebibliography}

\end{document}


\maketitle

\section{Experimental methods}

Sample preparation was accomplished in an ultrahigh vacuum (UHV) system with a base pressure of $p<2\times10^{-10}$\,mbar. Ir(111) was cleaned using cycles of
1\,kV $\mathrm{Ar^+}$-sputtering and subsequent flash annealing to 1520\,K. Graphene (Gr) was grown by ethylene exposure to saturation, subsequent flash annealing to 1470\,K and a final exposure to 800\,L ethylene at 1370\,K. The quality of the closed single crystal Gr monolayer was checked by low energy electron diffraction (LEED) and scanning tunneling microscopy (STM).

Monolayer H-\nbs{} was grown on Gr/Ir(111) by reactive molecular beam epitaxy (MBE). The substrate was exposed to an Nb flux of $\mathrm{5.8\times10^{15}}$\,atoms/$\mathrm{m^2s}$ from an e-beam evaporator in an elemental sulfur (S) background pressure of $p=1\times 10^{-8}$\,mbar created by a pyrite (FeS$_2$) filled Knudsen cell. The growth was conducted for 660\,s at 300\,K substrate temperature. Subsequently, the sample was annealed to 800\,K to improve the layer quality~\cite{Hall18}. In order to maximize the monolayer coverage, the island seeds were extended to final size by additional growth at 800\,K for 660\,s.

After synthesis, the H-\nbs{} layer was checked using LEED. Subsequently, the sample was transferred in UHV to the connected UHV bath cryostat STM chamber for investigation. The temperature $T_\mathrm{s}$ of STM or scanning tunneling spectroscopy (STS) investigation is specified in each figure and was either 0.4\,K using a He$^3$ cycle, 1.7\,K when pumping on He$^4$, 4\,K using He$^4$ cooling without pumping, or even higher than 4\,K by using an internal heater. Dependence of the STS features on magnetic field was checked by a superconducting magnet creating fields of up to 9\,T normal to the sample surface.

Both, constant-height and constant-current modes, were used to measure topography and d$I$/d$V$ maps. d$I$/d$V$ spectra were recorded only at constant height. Constant-current STM topographs and constant-current d$I$/d$V$ maps were recorded with sample bias $V_\mathrm{s}$ and tunneling current $I_\mathrm{t}$ specified in corresponding figure captions. Constant-height d$I$/d$V$ spectra and d$I$/d$V$ maps were recorded with stabilization bias $V_\mathrm{stab}$ and stabilization current $I_\mathrm{stab}$ using a lock-in amplifier with a modulation frequency $f_\mathrm{mod}$ and modulation voltage $V_\mathrm{mod}$ specified in corresponding captions. In case that for a constant-height d$I$/d$V$ map the sample bias during measurement does not coincide with $V_\mathrm{stab}$, the sample bias $V_\mathrm{s}$ is specified additionally. When needed, a voltage divider was applied to improve resolution. To ensure a reproducible and flat tip density of states (DOS) for the STS measurements, Au-covered W tips were used and calibrated beforehand using the surface state of Au(111)~\cite{Kaiser86,Everson91}.
Details on STM image processing are given in Figure~\ref{si_fig_cdw_fft}.

\section{Low-energy electron diffraction (LEED)}

\begin{figure}[H]
	\centering
	\includegraphics[width=0.5\textwidth]{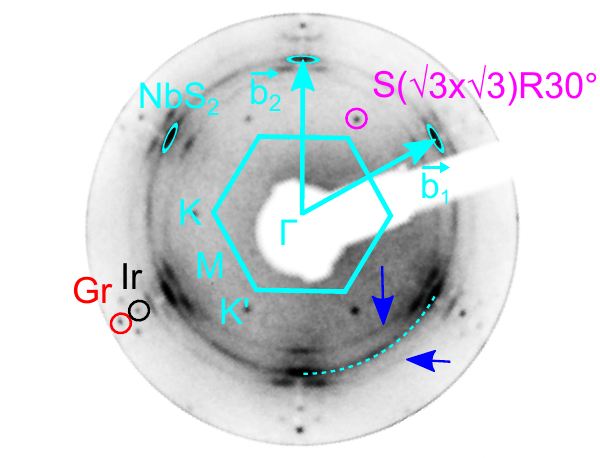}
	\caption{100\,eV microchannel plate LEED pattern. First order reflections of \nbs~are marked in turquoise, of Gr in red, of Ir in black, and of S intercalated between Gr and Ir in pink. Two faint moir\'e satellite rings are highlighted by blue arrows. Primitive reciprocal space translations $\vec b_1$ and $\vec b_2$ are indicated.}
	\label{si_fig_leed}
\end{figure}

The LEED pattern corresponding to the STM topograph of Figure~1a displays first order \nbs~intensity as superposition of (i) prominent elongated spots (several encircled turquoise) reasonably aligned with Gr (encircled red) and Ir (encircled black) and (ii) a diffraction ring due to randomly oriented islands (dashed turquoise segment). Apparently, most islands are aligned with small angular scatter to Gr/Ir(111), while some display random orientation. Additionally, faint off-centered rings are visible (two highlighted by blue arrows). These rings belong to \nbs, but are each shifted by one moir\'e periodicity of Gr. S intercalation between Gr and Ir gives rise to a ($\sqrt{3} \times \sqrt{3}$)R30$^\circ$ structure with respect to Ir(111) and corresponding reflections, of which one first order reflection is encircled in pink.

\section{Local density of states of monolayer \nbs}

\begin{figure}[H]
	\centering
	\includegraphics[width=\textwidth]{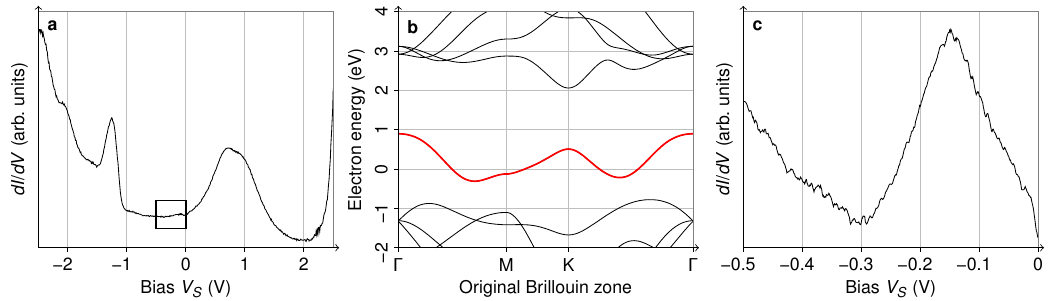}
	\caption{d$I$/d$V$ spectra of monolayer \nbs:
	(a)~Large-range d$I$/d$V$ spectrum from $-2.5$\,V to $+2.5$\,V.
	(b)~Calculated band structure of freestanding H-\nbs.
	(c)~d$I$/d$V$ spectrum of boxed voltage range in (a) with tip very close to sample. Spectra parameters:
	(a) $V_\mathrm{stab} = 2.5$\,V, $I_\mathrm{stab} = 1.0$\,nA, $V_\mathrm{mod} = 15$\,mV, $f_\mathrm{mod} = 797$\,Hz, $T_\mathrm{s} = 0.4$\,K;
	(c) $V_\mathrm{stab} = -0.5$\,V, $I_\mathrm{stab} = 0.5$\,nA, $V_\mathrm{mod} = 5$\,mV, $f_\mathrm{mod} = 1873$\,Hz, $T_\mathrm{s} = 0.4$\,K.}
	\label{si_fig_sts}
\end{figure}

To gain further insight into the electronic structure, differential conductance d$I$/d$V$ spectra were measured on \nbs, shown in Figure~\ref{si_fig_sts}. Figure~\ref{si_fig_sts}a displays a large-range constant-height d$I$/d$V$ spectrum which can be compared to the density functional theory (DFT) band structure in Figure~\ref{si_fig_sts}b. We note that (i) Van Hove singularities appear pronounced in d$I$/d$V$ spectra due to the large local DOS (LDOS) associated to them, and (ii) states with large parallel momentum $k_{||}$ are diminished or even suppressed in d$I$/d$V$ spectra, since a large $k_{||}$ is associated with a large decay constant $\kappa$, \emph{i.e.}, a rapid decay of the wave function into vacuum~\cite{Feenstra87,Tersoff83}.

The pronounced peak at $-1.25$\,V in Figure~\ref{si_fig_sts}a is attributed to the three occupied S $p$-bands with minima or maxima around $-1.25$\,V at the $\Gamma$-point in Figure~\ref{si_fig_sts}b. Additional maxima in the occupied states along the $\overline{\Gamma \textrm M}$ or the $\overline{\Gamma \textrm K}$ direction are hardly visible in the d$I$/d$V$ spectrum because of their larger $k_{||}$. The broad and intense maximum with its peak at about $+0.85$\,V in Figure~\ref{si_fig_sts}a is associated to the band maximum of the Nb $d_\textbf{z}$-type hole-like pocket at the $\Gamma$-point in Figure~\ref{si_fig_sts}b, though located at slightly lower energies as in the calculation. The steep rise in the d$I$/d$V$ spectrum at energies above about $+2.2$\,V is associated to the empty Nb $d$-bands with energies above 2\,eV in the calculated band structure. Figure~\ref{si_fig_sts}c displays an STS spectrum for the energy range from $-0.5$\,V to 0\,V (boxed in Figure~\ref{si_fig_sts}a). It is stabilized at $-0.5$\,V, \emph{i.e.}, at an energy with a low DOS as seen in Figure~\ref{si_fig_sts}a. To pick up the stabilization current of $I_\mathrm{t} = 0.5$\,nA the tip moves close to the surface and thus becomes sensitive to less pronounced features in the LDOS. The peak at $-0.15$\,V in the resulting spectrum can be interpreted as the Van Hove singularity associated with the toroidal minimum of the Nb $d$-band surrounding the $\Gamma$-point.

\section{Band structure of \nbs~near the $\Gamma$-point from quasi-particle interference}

\begin{figure}
	\centering
	\includegraphics[width=0.85\textwidth]{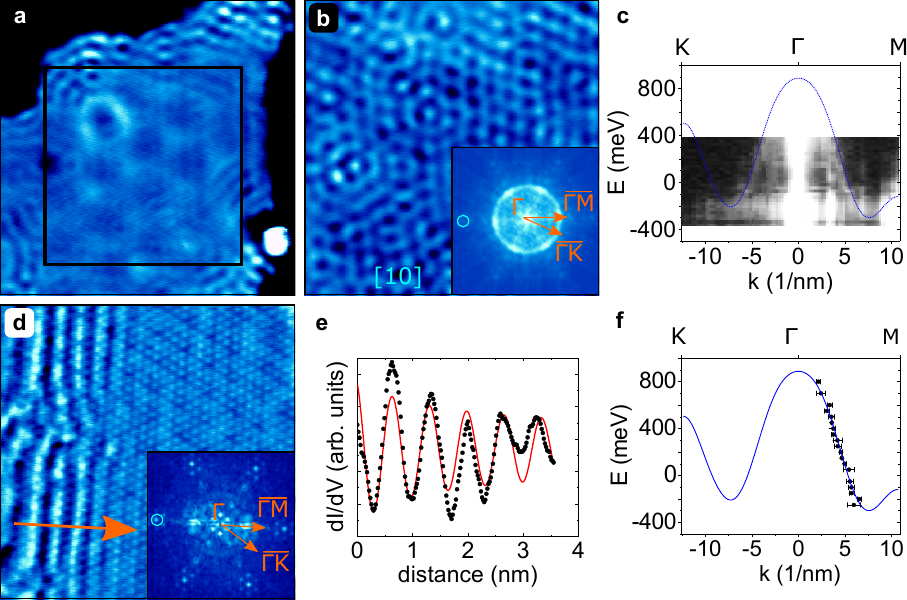}
	\caption{Quasi-particle interference in \nbs:
	(a)~Constant-current d$I$/d$V$ map.
	(b)~Constant-current d$I$/d$V$ map of the area inside the black square shown in (a). Inset displays the FFT of the d$I$/d$V$ map. An atomic lattice reflection is marked turquoise, the reciprocal space directions marked orange.
	(c)~FFT intensity of the d$I$/d$V$ maps along $\overline{\Gamma \textrm M}$ and $\overline{\Gamma \textrm K}$ as a function of energy $E = e V_\mathrm{s}$. Data extracted from a $\mathrm{200\times200}$ grid of constant-height d$I$/d$V$ spectra in an area of 9\,nm$\times$9\,nm on \nbs. To represent dispersion, experimental wave vectors are divided by a factor of two~\cite{Petersen98}. Superimposed as dotted-blue line is our DFT calculated dispersion of the H-\nbs~$d$-band.
	(d)~Constant-current d$I$/d$V$ map with the standing wave pattern due scattering at island edge. Direction of the wave vector indicated by an orange arrow. Inset is the FFT of the d$I$/d$V$ map.
	(e)~Line profile along the orange arrow in (d) (black dots) fitted by a Bessel function (red line)~\cite{Crommie93}.
	(f)~$E(k)$ dispersion relation extracted from the Bessel function fits and compared to the calculated band structure also displayed in (c).
	Image information:
	(a) size $\mathrm{13\,nm\times13}$\,nm, $V_\mathrm{s} = 200$\,mV, $I_\mathrm{t} = 1.0$\,nA, $V_\mathrm{mod} = 20$\,mV, $f_\mathrm{mod} = 797$\,Hz, $T_\mathrm{s} = 0.4$\,K;
	(b) size $\mathrm{9\,nm\times9}$\,nm, $V_\mathrm{s} = 30$\,mV, $I_\mathrm{t} = 0.8$\,nA, $V_\mathrm{mod} = 7$\,mV, $f_\mathrm{mod} = 1890$\,Hz, $T_\mathrm{s} = 0.4$\,K;
	(d) size $\mathrm{10\,nm\times10}$\,nm, $V_\mathrm{s} = 150$\,mV, $I_\mathrm{t} = 0.3$\,nA, $V_\mathrm{mod} = 20$\,mV, $f_\mathrm{mod} = 1890$\,Hz, $T_\mathrm{s} = 1.7$\,K.}
	\label{si_fig_standing_waves}
\end{figure}

Besides the charge density wave (CDW) superstructure, another electronic feature cannot be overlooked in \nbs~monolayer islands. The 100\,mV d$I$/d$V$ map in Figure~\ref{si_fig_standing_waves}a displays standing wave patterns at the \nbs~island edges originating from quasi-particle interference (QPI) of electron waves. Zooming into the island, Figure~\ref{si_fig_standing_waves}b shows a constant-current d$I$/d$V$ map recorded at 30\,mV. At this bias voltage damping of the QPI is weak and the interference pattern is spread out over the whole island. The inset with the fast Fourier transform (FFT) exhibits a ring-like feature, which shows enhanced intensity in the $\overline{\Gamma \textrm M}$ direction. QPI at 30\,meV is thus close to isotropic in wave vector, but anisotropic in scattering intensity.
The QPI pattern is used to extract the dispersion~\cite{Crommie93,Hasegawa93,Hormandinger94} of the $d$-band crossing the Fermi level, discussed in Figure~\ref{si_fig_sts}b.

In Figure~\ref{si_fig_standing_waves}c the FFT intensity profiles along the high-symmetry directions in $k$~space are plotted as function of energy. Superimposed to the dispersing feature in the data is the DFT calculated band as dotted-blue line. The bright cut-off toward larger $k$~values agrees reasonably with the calculation.

The dispersion is also determined by analysis of the real space periodicity of the standing waves at \nbs~island edges. Following the approach of Crommie \emph{et al.}~\cite{Crommie93}, the standing wave pattern resulting from backscattering at a straight island edge is fitted after proper background subtraction through $\mathrm dI/\mathrm dV [V_\mathrm{s},x]=L_0[1-J_0(2kx+\phi)]$. Here $J_0$ is the zeroth order Bessel function, $L_0=m^*/(\pi\hbar^2)$ with $m^*$ being the effective mass, $\phi$ is a phase constant, $x$ the distance from the step edge and $k$ is the wave vector related to the electron energy $E = e V_\mathrm{s}$. Figure~\ref{si_fig_standing_waves}e exemplifies our approach for a profile (black dots) taken along the orange arrow in the 150\,mV d$I$/d$V$ map shown in Figure~\ref{si_fig_standing_waves}e. The fit is shown as thin red line and yields the $k$~vector for $E = 150$\,meV. Figure~\ref{si_fig_standing_waves}f presents our analysis in the energy range from $-250$\,meV to 800\,meV (black dots), which compares favorably with our DFT calculated dispersion shown as blue line. 

\section{Details on FFT filtering of the d$I$/d$V$ maps}

\begin{figure}[H]
	\centering
	\includegraphics[width=0.85\textwidth]{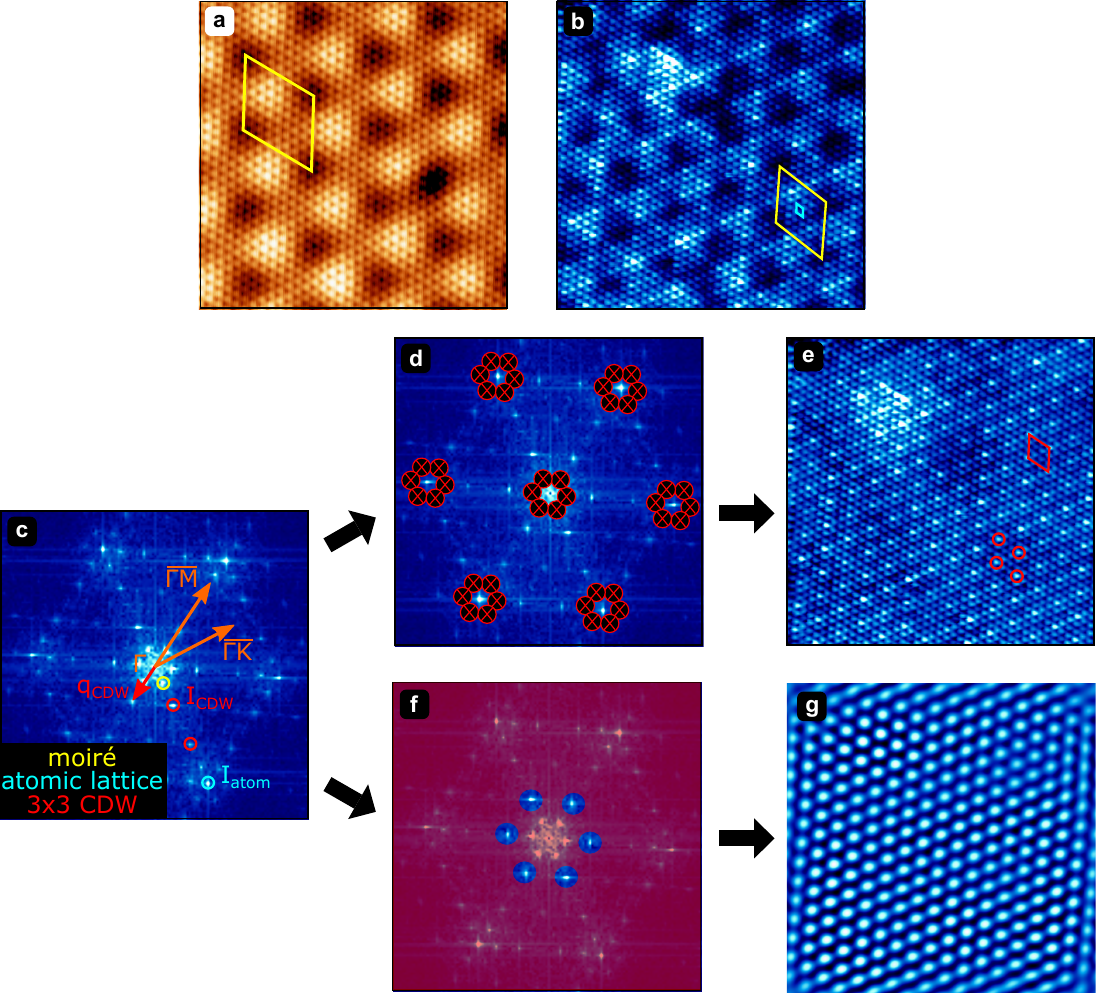}
	\caption{(a)~Atomically resolved constant-current STM topograph of monolayer \nbs. The Gr/Ir(111) moir\'e unit cell is indicated as yellow diamond.
	(b)~Constant-height d$I$/d$V$ map. The Gr/Ir(111) moir\'e and the \nbs~unit cell are indicated as yellow and turquoise diamonds, respectively.
	(c)~FFT of (b). Spots corresponding to the Gr/Ir(111) moir\'e, the atomic \nbs~lattice, and the $3\times3$ CDW superstructure are encircled yellow, turquoise, and red, respectively. A $\overline{\Gamma \textrm M}$- and a $\overline{\Gamma \textrm K}$-direction are indicated by orange arrows.
	(d)~The application of bandstop filtering to remove the moir\'e is visualized.
	(e)~Back transformed moir\'e bandstop filtered d$I$/d$V$ map. Red circles and rhomboid in (e) highlight $3\times3$ CDW superstructure.
	(f)~The application of bandstop filtering to remove all, but the $3\times3$ CDW superstructure spots is shown.
	(g)~Back transformed bandstop filtered image of (f) leaving only $3\times3$ periodicity in real space.
	Images:
	(a)~size $\mathrm{10\,nm \times 10\,nm}$, $V_\mathrm{s} = 100$\,mV, $I_{t} = 0.7$\,nA, $T_\mathrm{s} = 1.7$\,K;
	(b)~size $\mathrm{12\,nm\times12}$\,nm, $V_\mathrm{s} = - 15$\,mV, $V_\mathrm{stab} = 300$\,mV, $I_\mathrm{stab} =5$\,nA, $V_\mathrm{mod} = 5$\,mV, $f_\mathrm{mod} = 1890$\,Hz, $T_\mathrm{s} = 4$\,K.}
	\label{si_fig_cdw_fft}
\end{figure}

In the atomically resolved STM topograph of Figure~\ref{si_fig_cdw_fft}a the Gr/Ir(111) moir\'e (yellow diamond) can be recognized being superimposed on the atomically resolved \nbs~lattice. The lack of an own moir\'e between the \nbs~monolayer and Gr indicates a very weak interaction between the two materials.

The FFT of the $-15$\,meV constant-height d$I$/d$V$ map shown in Figure~\ref{si_fig_cdw_fft}b is presented as Figure~\ref{si_fig_cdw_fft}c. The red encircled spots at 1/3 and 2/3 of the distance between the $\Gamma$-point and the first order \nbs~lattice spots (encircled turquoise) are indicative of a $3\times3$ superstructure. It is obvious that the three equivalent wave vectors related to the superstructure are oriented along the $\overline{\Gamma \textrm M}$-directions. For better visualization of the $3\times3$ superstructure in real space, the moir\'e spots are bandstop filtered as shown in the Figure~\ref{si_fig_cdw_fft}d. Upon back transformation, the $3\times3$ superstructure highlighted by red circles and a rhomboid becomes obvious in Figure~\ref{si_fig_cdw_fft}e. Figure~\ref{si_fig_cdw_fft}e is identical with Figure~2a of the main text. The same procedure was implemented for Figure~2b,\,c of the main text.

To visualize the CDW without background disturbances all FFT spots except of the CDW spots can be bandstop filtered, as demonstrated in Figure~\ref{si_fig_cdw_fft}f. Upon back transformation only the CDW spots are visible in Figure~\ref{si_fig_cdw_fft}g. This technique was used to obtain the maps of Figure~2d in the main text.

\section{Doping of Gr underlayer effect on \nbs~CDW}

\begin{figure}[H]
	\centering
	\includegraphics[width=0.85\textwidth]{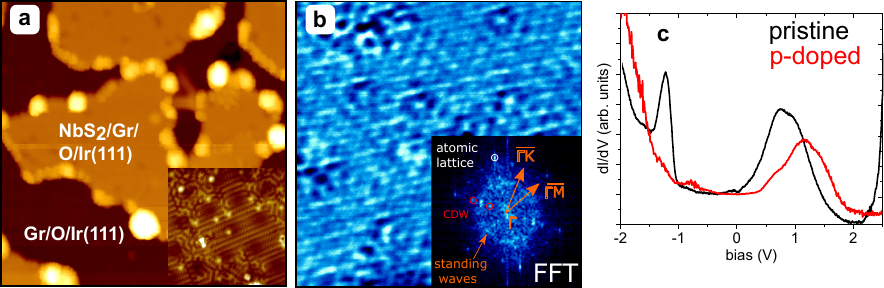}
	\caption{(a) Constant-current STM image of \nbs~on O-intercalated Gr on Ir(111). Inset: High-resolution topograph of Gr between \nbs~islands with stripes reflecting the ($2 \times 1$) adsorption pattern of atomic O on Ir(111) through the Gr layer~\cite{Galera16}.
	(b)~Constant-current d$I$/d$V$ map. Inset: FFT of (b) with CDW spots encircled red.
	(c)~Large range d$I$/d$V$ spectra of \nbs~on O-intercalated Gr/Ir(111) (red) compared to non-intercalated, pristine \nbs~(black) as in Figure~\ref{si_fig_sts}a.
	Image information:
	(a)~size $\mathrm{58\,nm\times58}$\,nm, $V_\mathrm{s} = 4$\,V, $I_\mathrm{t} = 0.1$\,nA, $T_\mathrm{s} = 1.7$\,K;
	(b)~size $\mathrm{10\,nm\times10}$\,nm, $V_\mathrm{s} = -200$\,mV, $I_\mathrm{t} = 0.7$\,nA, $V_\mathrm{mod} = 7$\,mV, $f_\mathrm{mod} = 1890$\,Hz, $T_\mathrm{s} = 10$\,K.
	Spectra information:
	(c)~red: $V_\mathrm{stab} = 2.5$\,V, $I_\mathrm{stab} = 1.0$\,nA, $V_\mathrm{mod} = 15$\,mV, $f_\mathrm{mod} = 797$\,Hz, $T_\mathrm{s} = 1.7$\,K: black: $V_\mathrm{stab} = 2.5$\,V, $I_\mathrm{stab} = 1.0$\,nA, $V_\mathrm{mod} = 15$\,mV, $f_\mathrm{mod} = 797$\,Hz, $T_\mathrm{s} = 0.4$\,K.}
	\label{si_fig_doping}
\end{figure}

Figure~\ref{si_fig_doping}a displays an STM topograph with \nbs~islands on O-intercalated Gr/Ir(111). Details of the intercalation method are described elsewhere~\cite{vanEfferen22}. From Figure~\ref{si_fig_doping}b it is obvious that the ($ 3 \times 3$) CDW superstructure is present. Figure~\ref{si_fig_doping}c compares the d$I$/d$V$ spectra of \nbs~on O-intercalated Gr/Ir(111) with the pristine case. The overall d$I$/d$V$ features of p-doped \nbs~(red curve) are shifted toward positive energy in respect to the pristine case (black curve), in agreement with p-doping.

\section{Real space visualization of inelastic d$I$/d$V$ features}

\begin{figure}[H]
	\centering
	\includegraphics[width=0.85\textwidth]{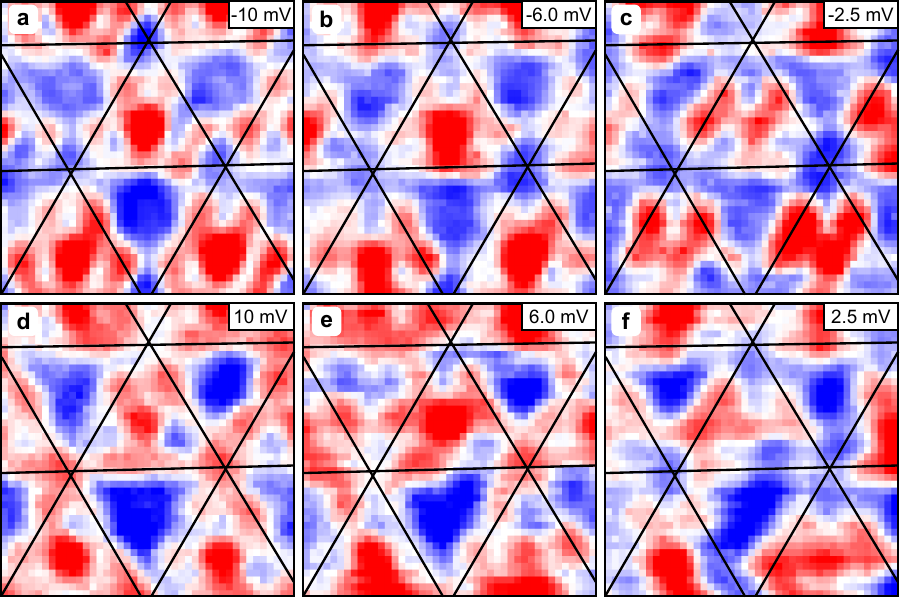}
	\caption{Constant-height d$I$/d$V$ maps at bias voltages of $-10$\,mV in (a), $-6$\,mV in (b), $-2.5$\,mV in (c), $+10$\,mV in (d), $+6$\,mV in (e), $+2.5$\,mV in (f). The energies are selected according to the energy positions of the inelastic excitations displayed in Figure~3b,\,c of the main text. The black lines indicate the diamonds of the ($3 \times 3$) CDW superstructure, which are additionally segmented by them into two triangular areas.
	Image information:
	(a--f)~size $\mathrm{2\,nm\times2}$\,nm, $V_\mathrm{stab} = 100$\,mV, $I_\mathrm{stab} = 4.7$\,nA, $V_\mathrm{mod} = 0.5$\,mV, $f_\mathrm{mod} = 311$\,Hz, $T_\mathrm{s} = 0.4$\,K}
	\label{si_fig_didvmaps}
\end{figure}

The six constant-height d$I$/d$V$ maps in Figure~\ref{si_fig_didvmaps} are taken in the white box of Figure~3a and at the energies indicated by the dashed lines of Figure~3b of the main text. Although the interpretation of the local variation of the d$I$/d$V$-intensity is not straight forward and may certainly be affected by details of the tip apex, it is obvious that the intensity variation in all maps reflects the ($3 \times 3$) CDW periodicity. The down-pointing triangles generally possess lower intensity and the up-pointing triangles higher intensity, the later varying in lateral intensity distribution as a function of energy.

\section{Magnetic field dependence of the inelastic excitations}

\begin{figure}[H]
	\centering
	\includegraphics[width=0.4\textwidth]{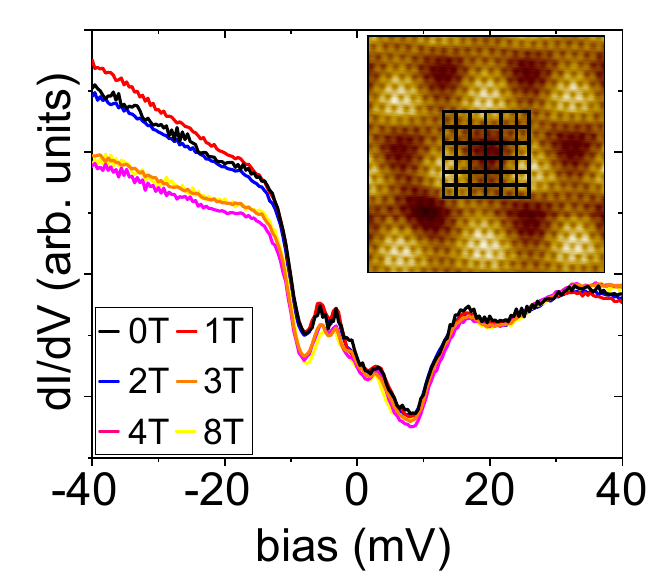}
	\caption{Average spectra at different magnetic field applied normal to the surface, as indicated. Each spectrum in a set of 49 spectra for one magnetic field is taken at a location defined by the grid in the inset. Inset: Constant-current STM topograph of \nbs.
	Image information:
	Inset: size $\mathrm{7\,nm\times7}$\,nm, $V_\mathrm{s} = 100$\,mV, $I_\mathrm{t} = 0.7$\,nA, $T_\mathrm{s} = 1.7$\,K. Spectra information:
	$V_\mathrm{stab} = 40$\,mV, $I_\mathrm{stab} = 0.7$\,nA, $V_\mathrm{mod} = 0.5$\,mV,
	$f_\mathrm{mod} = 797$\,Hz, $T_\mathrm{s} = 0.4$\,K.}
	\label{si_fig_magfield}
\end{figure}

Figure~\ref{si_fig_magfield} shows a data set different from the one represented in Figure~3 of the main manuscript. Each point spectrum shown is an average of $49$ d$I$/d$V$ spectra taken on a grid defined by the inset of Figure~\ref{si_fig_magfield}. Again, low energy features within the trough gap are well visible. No change of the average spectra is found as a function of external magnetic field up to 8\,T. The somewhat larger d$I$/d$V$ intensity at negative voltages and fields of 3\,T, 4\,T and 8\,T is presumably a drift effect.

\section{Computational methods}

Simulations of metals require a so-called smearing factor for stabilization of the calculations. Here, we use Marzari-Vanderbilt cold smearing~\cite{Marzari1999}. Compared to the Fermi-Dirac distribution with a finite temperature $T$, this cold smearing has the advantage that the low-temperature behavior (most experimental measurements here were performed at 4\,K) can be estimated from electronic structure calculations at larger broadening and therefore sparser Brillouin-zone sampling. In any case, even using the Fermi-Dirac distribution as the smearing function still disregards thermal motion of the nuclei and therefore overestimates the critical temperature. Furthermore, effects such as hybridization with substrates can have a similar smearing-like influence on lattice instabilities as electronic temperature~\cite{Hall19}. In some figures, we show results as a function of the smearing $\sigma$ to illustrate how stable the results are and as an indication for the influence of temperature. Note that the mentioned smearing values are only used for the structural relaxation, not for the electronic and phononic DOS.

For our downfolding, we consider an effectively noninteracting model with a linearized electron-lattice coupling. Its free energy as a function of atomic displacements reads $F(\vec u) = E \super{el} (\vec u) - T S \super{el} (\vec u) + E \super{lat} (\vec u) + E \super{dc} (\vec u)$ with the total single-electron energy $E \super{el}$ of the linearized low-energy Kohn-Sham Hamiltonian $H_0 \super{el} + \vec u \vec d$, the corresponding generalized entropy $S \super{el}$, as well as the quadratic lattice term $E \super{lat}$ and the linear double-counting term $E \super{dc}$, chosen such that the second and first order of the free energy match DFT and density functional perturbation theory (DFPT) for the undistorted system.

The calculations for the undistorted system are done with \textsc{Quantum ESPRESSO}~\cite{Giannozzi2009, Giannozzi2017, Giannozzi2020}. We apply the PBE functional~\cite{Perdew1996} and normconserving pseudopotentials from \textsc{PseudoDojo}~\cite{Hamann2013, vanSetten2018} at an energy cutoff of 100\,Ry. A Marzari-Vanderbilt cold smearing~\cite{Marzari1999} of $\sigma_0 = 20$\,mRy is combined with uniform $12 \times 12$ $\vec k$ and $\vec q$ meshes including $\Gamma$. When going to lower values $\sigma$ on the model level, the number of $\vec k$~points per dimension is scaled by a factor of $\lceil \sigma_0 / \sigma \rceil$.
Phonon dispersions of the undistorted system for low smearings have been obtained in a computationally efficient way from the data for the highest smearing using the method of Ref.~\citenum{Calandra2010}, which has proven to yield excellent results for TaS\s2~\cite{Berges2022}. Here, we generalize this method with respect to distorted structures on supercells. To separate periodic images of the monolayer, we choose a unit-cell height $c = 2$\,nm and truncate the Coulomb interaction in this direction~\cite{Sohier2017}. The relaxed lattice constant $a = 0.335$\,nm is close to the experimental value. The downfolding to the low-energy model in the localized representation of atomic displacements and Wannier orbitals (Nb $d_{z^2}$, $d_{x^2 - y^2}$, and $d_{x y}$) is accomplished using \textsc{Wannier90}~\cite{Pizzi2020} and the EPW code~\cite{Giustino2007, Noffsinger2010, Ponce2016}.

Note that including spin-orbit coupling into the calculation leads to a band splitting and thus increases the number of peaks in the DOS. Nevertheless, this does not lead to an explanation of the experimentally observed d$I$/d$V$ spectra.

\section{Generalized free energy}

\begin{figure}[H]
	\centering
	\includegraphics{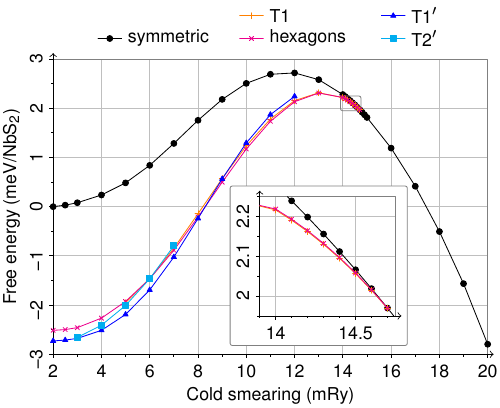}
	\caption{Free energy for different stable (CDW) structures as a function of cold smearing. Inset: Close-up of the region of the phase transition.}
	\label{fig:free_energy}
\end{figure}

Figure~\ref{fig:free_energy} shows the generalized free energy for the symmetric and the four distorted phases as a function of cold smearing $\sigma$. Here, all colored points correspond to fully relaxed structures. The structure could always be unambiguously classified as one of the four structures shown in the main text, even though the absolute and relative displacements change with the smearing. Below the critical smearing $\sigma \sub{CDW} \approx 14.7$\,mRy, the energy gain from the lattice distortion continuously increases up to about $2.8$\,mRy/\nbs{}. Here, the energy difference between the different CDW structures is very small, most likely smaller than the expected accuracy of our theoretical approach. Thus, Figure~\ref{fig:free_energy} should not be considered the final answer to the question of which CDW structure is observed and we consider all structures in the following. Not all structures are stable at all smearings; ``hexagons'' and T1 are favored directly below the CDW transition, T1$'$ and T2$'$ for smaller smearings. Only T1 is found for the whole smearing range considered.

\section{Four possible CDW phases}

\begin{figure}
	\includegraphics[width=\linewidth]{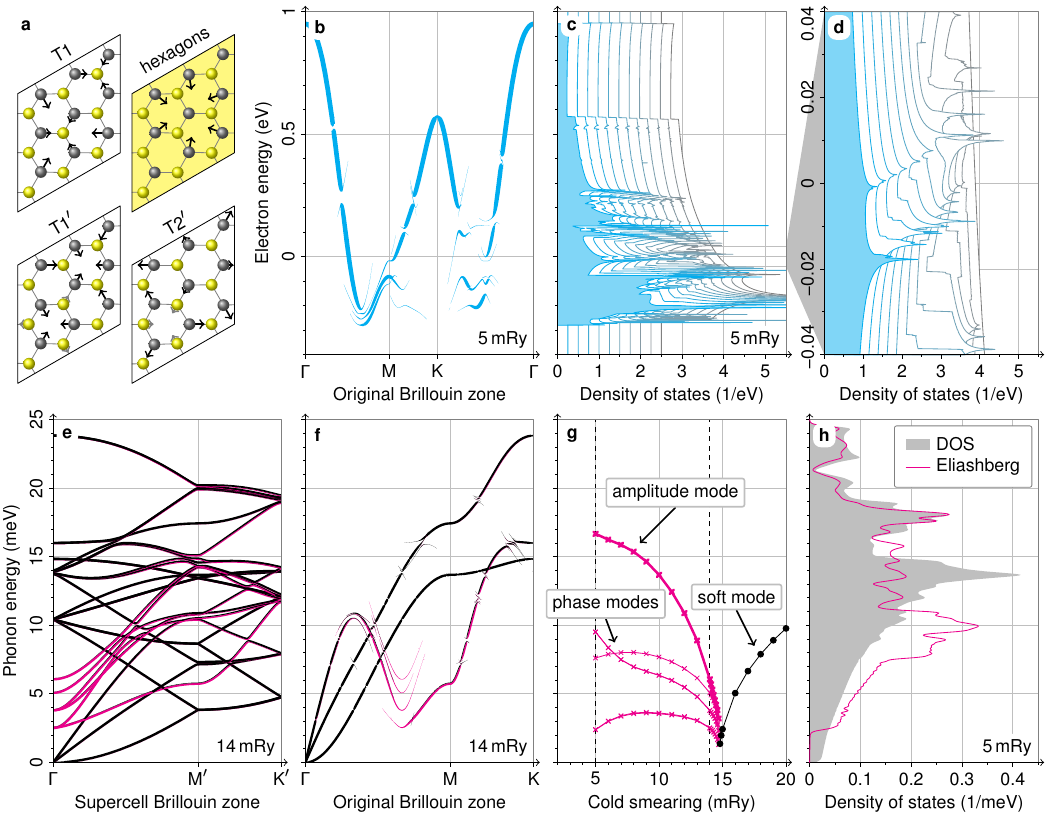}
	\caption{Electronic and phononic properties in the ``hexagons'' CDW phase, cf.\@ T1 in the main text.}
	\label{fig:tonb}
\end{figure}

\begin{figure}
	\includegraphics[width=\linewidth]{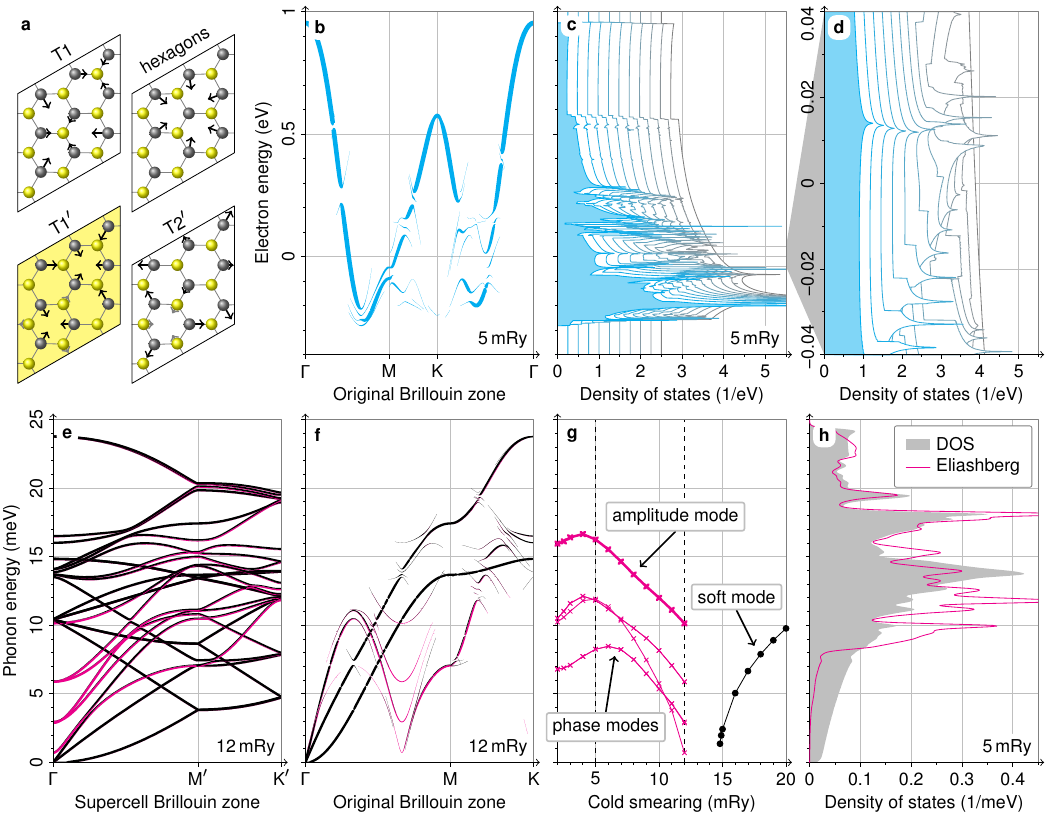}
	\caption{Electronic and phononic properties in the T1$'$ CDW phase, cf.\@ T1 in the main text.}
	\label{fig:togap}
\end{figure}

\begin{figure}
	\includegraphics[width=\linewidth]{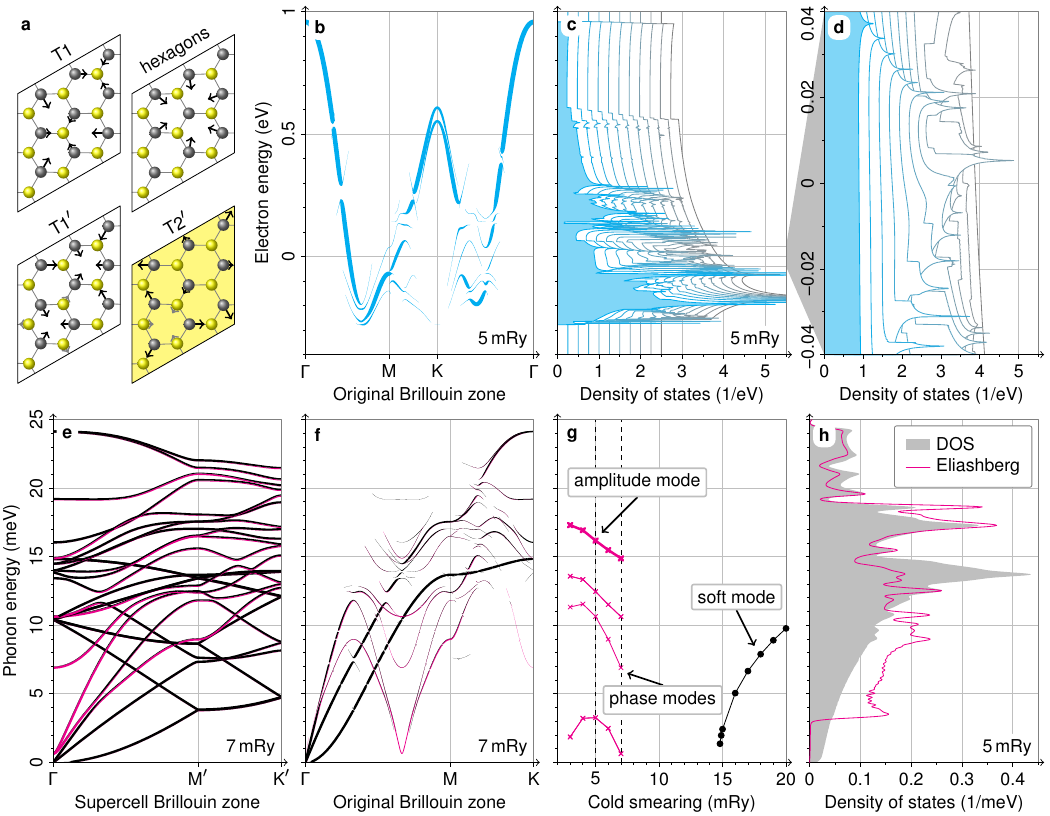}
	\caption{Electronic and phononic properties in the T2$'$ CDW phase, cf.\@ T1 in the main text.}
	\label{fig:fromgap}
\end{figure}

In our calculations, four different CDW phases were stabilized, which we denote as T1, ``hexagons'', T1$'$, and T2$'$~\cite{Guster2019}. In the main text, we have shown detailed information about the T1 phase. Here, the corresponding results for the other three phases are shown in Figures~\ref{fig:tonb}, \ref{fig:togap} and \ref{fig:fromgap}.

The electronic band structure and DOS are shown in panels (b--d). For panels (c--d), different displacement amplitudes (with respect to the undistorted structure) are shown, with gray corresponding to the undistorted structure and blue to the fully distorted structure whose band structure is shown in (b). In all cases, the reduction of the symmetry in the distorted phases leads to the appearance of additional bands and additional Van Hove singularities. The position of these Van Hove singularities depends on the displacement, often approximately linearly, and the magnitude of the changes is on the 100 meV scale. Thus, although it is possible to interpret peaks in the experimental STS as Van Hove singularities, fine-tuned parameters are needed to place these peaks at the desired position close to the Fermi level.

Panels (e) and (f) show the phonon dispersions in the supercell and original Brillouin zone, respectively. The magenta marking shows to what extent these phonons correspond to the unstable phonon modes in the undistorted structure. To be more precise, the absolute value squared of the scalar product of the respective displacements determines the fraction of the line that is color magenta. For each plot, the smearing is listed in the bottom right. Panel (g) shows all smearings where the structure can be stabilized and the energies of the phase and amplitude phonons at the $\Gamma$~point in the supercell as a function of smearing. Figure~\ref{fig:fromgap}(g) in particular shows that the vanishing energy of one of the phonon modes denotes the end of the stable region in parameter space: at the transition point, a local minimum in the free energy becomes a local maximum. Although their details differ, all four structures have phase and amplitude modes at very similar energy scales of approximately 10 meV. Moving over to panel (h), we show the phononic DOS and the Eliashberg function $\alpha^2 F(\omega)$. The amplitude and phase modes at $\Gamma$ are hardy visible in the phononic DOS, which is dominated by acoustic phonons.

For the formation of polaronic excitations, we need to know both the frequencies at which there are phonons and how strongly these phonons are coupled to the electrons. This can be quantified using the Eliashberg spectral function $\alpha^2 F(\omega)$. The electron-phonon coupling appears squared in this expression since the electron needs to emit and absorb a phonon. The Eliashberg spectral function is shown in panel (h) of the figures. The Eliashberg spectral function has a clear onset at the energy corresponding to the lowest phase mode. This shows that the modes corresponding to the longitudinal-acoustic modes at $\vec q = 2/3\,\overline{\Gamma \textrm M}$ in the undistorted state still dominate the coupling in presence of the CDW, due to their large electron-phonon matrix elements~\cite{Lian2023}. On the other hand, the phonon DOS itself has contributions all the way down to zero frequency, coming from the acoustic branches, but these are weakly coupled to the electrons and irrelevant for the formation of combined electron-phonon excitations.

\bibliography{ms}